\documentclass[aps,prd,twocolumn,superscriptaddress,groupedaddress,nofootinbib]{revtex4}
\usepackage{graphicx}  % needed for figures
\usepackage{dcolumn}   % needed for some tables
\usepackage{bm}        % for math
\usepackage{amssymb}
\usepackage{amstext}% for math
\usepackage{amsmath}
\usepackage{dsfont}
\usepackage{xcolor}
\usepackage{amsthm}
\usepackage{enumitem}
\usepackage{tensor}
\usepackage{float}
\usepackage{array}
\usepackage{tikz}  
\usepackage[colorlinks]{hyperref}
\hypersetup{
     breaklinks=true,
    pdfstartview={FitH},    % fits the width of the page to the window
    colorlinks=true,       % false: boxed links; true: colored links
    linkcolor=blue,          % color of internal links
    citecolor=red,        % color of links to bibliography
    filecolor=magenta,      % color of file links
    urlcolor=blue,           % color of external links
    anchorcolor=green,      % Color for anchor text
    linktocpage=true
}

\newcommand{\be}{\begin{equation}}\newcommand{\ee}{\end{equation}}
\newcommand{\bea}{\begin{eqnarray}}\newcommand{\eea}{\end{eqnarray}}
\newcommand{\brr}{\begin{array}}\newcommand{\err}{\end{array}}
\newcommand{\bit}{\begin{itemize}}\newcommand{\eit}{\end{itemize}}
\newcommand{\ben}{\begin{enumerate}}\newcommand{\een}{\end{enumerate}}

\newcommand{\ba}{\begin{array}}
\newcommand{\ea}{\end{array}}

\definecolor{darkred}{rgb}{.8,0,0}
\definecolor{darkblue}{rgb}{0,0,0.7}

\def\T{{_{T}}}

\def\1{{_{1}}}\def\2{{_{2}}}

\def\noHe0{:\;\!\!\;\!\!:H_e(0):\;\!\!\;\!\!:}
\def\noHm0{:\;\!\!\;\!\!:H_\mu(0):\;\!\!\;\!\!:}

\begin{document}
%%%%%%%%%%%%%%%%%%%%%%%%%%%%%%%%%%%%%%%%%%%%%%%%%%%%%%%%%%%%%%%
\title{Direct product geometries in $n$-dimensional linear and quadratic gravity}

%%%%%%%%%%%%%%%%%%%%%%%%%%%%%%%%%%%%%%%%%%%%%%%%%%%%%%%%%%%%%%%
\author{Petr Jizba}
\email{p.jizba@fjfi.cvut.cz}
\affiliation{FNSPE,
Czech Technical University in Prague, B\v{r}ehov\'{a} 7, 115 19, Prague, Czech Republic}
\author{Tereza Lehe\v{c}kov\'{a}}
\email{lehecter@fjfi.cvut.cz}
\affiliation{FNSPE,
Czech Technical University in Prague, B\v{r}ehov\'{a} 7, 115 19, Prague, Czech Republic}
\date{\today}
\begin{abstract}
%%%%%%%%%%%%%%%%%%%%%%%%%%%%%%%%%%%%
Direct-product spacetimes offer a tractable route to exact solutions in gravity theories beyond general relativity. We study $n$-dimensional direct-product spacetimes of $(m_1+m_2)$, $(2+m)$, and $(2+\ldots+2)$ type in linear gravity  and general pure quadratic-curvature gravity, allowing the two-dimensional subspaces to have nonconstant curvature. We classify the resulting solution spaces and determine how the allowed geometries depend on the gravitational theory, the dimension, and the curvature couplings. Metrics containing two-dimensional part(s) of nonconstant curvature are governed either by algebraic conditions on a curvature profile,  or, in one special
class of quadratic gravity, by a differential equation. Relations
governing metrics with only constant-curvature $m$-dimensional
subspaces are purely algebraic. {We compare the two considered theories and study relations between their solution sets.} We further examine conformal
relations, generalized Petrov types,
and compatibility with electromagnetic sources. 
Four- and six-dimensional cases are analyzed in detail, and so is the Weyl-squared action (that reduces to Weyl conformal gravity in $n=4$), as a particular example of quadratic gravity couplings. Explicit analytic solutions illustrating the
resulting structure are presented whenever possible.

%%%%%%%%%%%%%%%%%%%%%%%%%%%%%%%%%
\end{abstract}

\vskip -1.0 truecm

\maketitle

%%%%%%%%%%%%%%%%%%%%%%%%%%%%%%
\section{Introduction}
\label{intro}
%%%%%%%%%%%%%%%%%%%%%%%%%%%%%%%%

This work builds in part on and substantially extends our previous analysis of  $(2+2)$ direct-product (DP) spacetimes (STs) in four-dimensional Weyl conformal gravity (WCG)~\cite{previous}. Here, we significantly broaden the scope of that analysis with respect to the class of DP geometries considered, the ST dimension, and the gravity theories investigated. A fully general analysis of DP STs, however, becomes technically challenging already for the $(3+1)$ DP structure. To keep the problem tractable, we restrict our attention to a class of geometries that is broad enough to remain physically interesting, while allowing the field equations (FEs) to be reduced to a manageable form: the metrics are constructed from one or more $m$-dimensional ($m$D) constant-curvature subspaces, possibly supplemented by general, otherwise unconstrained two-dimensional (2D) subspaces.  Such block-structured metrics are far from a mere technical convenience: DP geometries of this type arise, for instance, as near-horizon limits of black holes~\cite{Bertotti1959,Robinson1959,KunduriLucietti2013} and as backgrounds used in Kaluza--Klein compactifications and dimensional reductions~\cite{Maia1985,OverduinWesson1997}. Furthermore, they retain enough nontrivial geometric structure to probe the genuine content of higher-derivative gravity theories.

Throughout this paper, we restrict our attention to ST dimensions $n\geq4$. This is the natural range for the present analysis, both because the theories and DP models considered here become trivial or highly restricted for $n<4$, see Sec.~\ref{lowerdim}, and because the generalized Petrov classification used here is not available in lower dimensions, where the Weyl tensor vanishes identically. Our main aim is to determine how the allowed DP solutions depend on the gravitational theory, the ST dimension, and the couplings. In particular, we compare two settings.

The \emph{first} is linear gravity (LG), by which we mean Einstein gravity with a cosmological constant in arbitrary dimensions~\cite{EmparanReall2008}.  This provides the reference point against which the higher-curvature theories can be compared. The \emph{second} is pure quadratic gravity (QG), whose action contains curvature-squared terms but no Einstein--Hilbert or cosmological-constant terms. In its most general form, the quadratic-curvature action includes all independent curvature-squared scalars. A closely related theory that also includes the Einstein--Hilbert and cosmological terms is full quadratic gravity, sometimes referred to as Stelle's gravity, which was shown to be perturbatively renormalizable in~\cite{Stelle1977} and studied at the classical level in~\cite{Stelle1978}. Although a number of exact solutions are known in this theory~\cite{PravdaEtAl2017, QG2, QG3, QG4, QG5}, it is quite resistant to analytical treatment, and many of its solutions, such as the non-Schwarzschild black holes, were found and studied only numerically~\cite{LuPerkinsPopeStelle2015a,LuPerkinsPopeStelle2015b}. Here, by contrast, we consider the purely quadratic-curvature action without these lower-order terms, and restrict attention to DP geometries rather than spherical symmetry.
We show that this combination reduces the FEs to a form that admits an exact analytic treatment in arbitrary dimension, offering thus a complementary and fully analytic vantage point on the solution space of QG. 
As an example of a particular choice of couplings, we consider the Weyl-squared action, which in four dimensions reduces to WCG~\cite{Bach1921}. In higher dimensions this action is no longer conformally invariant. The analysis of a conformally invariant higher-dimensional generalization of WCG will be presented in a follow-up work.

The comparison between LG and QG is especially useful, as LG is by far the best-studied gravity theory in higher dimensions, and therefore provides a natural reference point~\cite{EmparanReall2008,Tangherlini1963,MyersPerry1986}. Such a comparison identifies which solution sectors are shared by both classes of theories and clarifies which properties of the DP geometries are genuinely tied to the higher-curvature terms rather than being shared with LG. It also allows us to examine whether solutions of QG that are not themselves LG solutions can nevertheless be related to LG solutions by Weyl transformations, i.e., whether they share the same causal structure. Moreover, since general relativity (the $n=4$ case of LG) is extremely well tested in the experimentally accessible regime, any viable extension of Einstein gravity should reproduce, or at least admit, the corresponding general-relativistic solutions in the appropriate limit.

DP geometries have a long history in Einstein gravity. They represent exact backgrounds in 4D Einstein--Maxwell theory~\cite{OrtaggioPodolsky2002,Kadlecova2009} and, more recently, in higher dimensions~\cite{Podolsky2012}, and their geometric and algebraic properties have been studied extensively in higher-dimensional gravity~\cite{Pravda2007,Pravda2013}.
Independently, exact solutions of QG have been investigated using Petrov-type classifications and methods based on Kundt and Kerr--Schild--Kundt ansatzes~\cite{PravdaEtAl2017,GursesSismanTekin2012}, but, to our knowledge, not within the systematic DP framework adopted here. Our aim is to go beyond these more specialized treatments by considering, within a common framework, several classes of DP geometries, namely those with $(m_1+m_2)$, $(2+m)$, and $(2+\ldots+2)$ structures, in arbitrary dimensions, and by comparing their solution spaces in LG and QG. The resulting classification separates solutions in which all DP subspaces have constant curvature from those in which a 2D subspace has genuinely nonconstant curvature. Among the theories considered here, the latter possibility is allowed only for certain combinations of the quadratic couplings, and is governed either by algebraic conditions on a curvature profile or, for one particular combination of the coupling constants, by a differential equation for the 2D curvature.
The special role of 4D also becomes apparent, {e.g., via} several restrictions that are unavoidable in higher dimensions disappear in 4D, This special role, together with the obvious physical relevance of four dimensions, is why we discuss the 4D case in detail in all three theories. In addition, we analyze the 6D case as a tractable higher-dimensional example.

The paper is organized as follows. In Section~\ref{def,id}, we establish our conventions and introduce the basic definitions and identities used
throughout the paper. In Section~\ref{LG}, we find solutions for STs with $(m_1+m_2)$, $(2+m)$, and $(2+\cdots+2)$ DP structures in the framework of LG. In Section~\ref{QG}, we introduce QG and derive its FEs for DP STs with the same structures as in Section~\ref{LG}. For each class of STs, we classify the resulting solutions across the space of QG theories and
examine their relation to the corresponding LG solutions. Finally, in Section~\ref{concl}, we summarize our results and discuss their implications. Additional technical and expanded material is collected in the six accompanying appendices. In Appendix~\ref{APetrov}, we review the higher-dimensional Petrov classification and apply it to the geometries discussed; in Appendix~\ref{ACEST}, we investigate conformal relations for $(m+1)$ geometries; in Appendices~\ref{Abox} and~\ref{Afullbox}, we present a number of particular solutions of the equation $\Box R + CR=0$ arising in Sec.~\ref{QG}; and in Appendix~\ref{AEM}, we study the incorporation of an electromagnetic (EM) field into some of our models. Lastly, in Appendix~\ref{Agraph}, we give a tabular summary of the solution spaces obtained in both theories, together with the relations between them, including the 4D, 6D, and Weyl-squared cases.

%%%%%%%%%%%%%%%%%%%%%%%%%%%%%%%%%%%%%%%%%%%%%%%%%%%%%%
\section{Important definitions and identities}
\label{def,id}
%%%%%%%%%%%%%%%%%%%%%%%%%%%%%%%%%%%%%%%%%%%%%%%%%%%%

Throughout the paper, we use the following conventions. The Riemann tensor is defined as
\begin{equation}
R^{\alpha}_{~\beta\mu\nu}
=
\partial_{\mu}\Gamma^{\alpha}_{~\beta\nu}
-
\partial_{\nu}\Gamma^{\alpha}_{~\beta\mu}
+
\Gamma^{\alpha}_{~\kappa\mu}\Gamma^{\kappa}_{~\beta\nu}
-
\Gamma^{\alpha}_{~\kappa\nu}\Gamma^{\kappa}_{~\beta\mu}\,,
\end{equation}
with the Ricci tensor and Ricci scalar given by
\begin{equation}
R_{\mu\nu}\ = \ R^{\alpha}_{~\mu\alpha\nu}\,,
\qquad
R\  = \ R^{\mu}_{~\mu}\, .
\end{equation}
We adopt the ``mostly plus'' metric signature $(-,+,\cdots,+)$ and use the notation
\begin{equation}
R_{\mu\nu}^2 \ \equiv \  R_{\mu\nu}R^{\mu\nu}\, ,
\qquad
R_{\mu\nu\kappa\lambda}^2
\ \equiv \ 
R_{\mu\nu\kappa\lambda}R^{\mu\nu\kappa\lambda}\, .
\end{equation}
Under a Weyl transformation, the transformed quantities are denoted by a
tilde. In particular,
\begin{equation}
\label{WTr}
g_{\mu\nu}
\ \rightarrow \ 
\tilde{g}_{\mu\nu}
\ = \ 
\Omega^2 g_{\mu\nu}
\ \equiv \ 
e^{2\omega}g_{\mu\nu}\, .
\end{equation}
The subject of this paper are DP STs, i.e., STs whose metrics can be written as a direct sum of metrics on independent subspaces of lower dimension, i.e. 
\begin{equation}
ds^2 \ = \  ds_0^2 \ + \  ds_1^2 \ + \  \cdots\, .
\end{equation}
By their very construction, many geometric objects, such as the Riemann tensor, Ricci tensor, and Ricci scalar, in DP STs can be split into sums of contributions associated with particular ST subspaces. We will be particularly interested in metrics containing one or more $m$D parts of constant curvature and/or general 2D parts.
Let us first recall some basic properties of the latter.

For a 2D metric $g_{ij}$ (or the 2D part of a
higher-dimensional metric), the Riemann tensor takes the form
\begin{equation}
\label{2DRiemann}
R_{ijkl} \ = \ \frac{1}{2}R\,(g_{ik}g_{jl} \ - \  g_{il}g_{jk})\, ,
\end{equation}
where the indices $i,j,k,l$ run only over the 2D
part~\cite{book2}. Contracting Eq.~(\ref{2DRiemann}) yields
\begin{equation}
\label{2DRiccit}
R_{ij} \ = \ \frac{1}{2}R\,g_{ij}\, ,
\end{equation}
for the Ricci tensor $R_{ij}$ and Ricci scalar $R$, and consequently
\begin{equation}
\label{2Dkvad}
R_{ijkl}^2 \ = \ R^2\, , \qquad R_{ij}^2 \ = \ \frac{R^2}{2}\, .
\end{equation}  
An important property of 2D metrics is their conformal
flatness, which allows the general Euclidean 2D
metric to be written as~\cite{Chern:1955}
\begin{equation}
\label{cf2DE}
ds^2_{(E2D)} \ = \ e^{2\omega(x,y)}\left(dx^2 \ + \ dy^2\right)\, ,
\end{equation}
with Ricci scalar
\begin{equation}
\label{ER}
R_E \ = \ -2e^{-2\omega(x,y)}\left(\frac{\partial^2 \omega(x,y)}{\partial x^2}
\ + \  \frac{\partial^2 \omega(x,y)}{\partial y^2}\right)\, .
\end{equation}
The Lorentzian counterpart may be written as~\cite{,Stephani:2003tm}
\begin{equation}
\label{cf2DL}
ds^2_{(L2D)} \ = \ e^{2\omega(t,r)}\left(-dt^2 \ + \  dr^2\right)\, ,
\end{equation}
with
\begin{equation}
\label{LR}
R_L \ = \ 2e^{-2\omega(t,r)}\left(\frac{\partial^2 \omega(t,r)}{\partial t^2}
- \frac{\partial^2 \omega(t,r)}{\partial r^2}\right)\, .
\end{equation}
For an $m$D metric $g_{ab}$ (or the $m$D part of a
higher-dimensional metric) of constant curvature, the Riemann tensor takes
the form
\begin{equation}
\label{mDccRiemann}
R_{abcd} \ = \ K\,(g_{ac}g_{bd} \ - \ g_{ad}g_{bc})\, ,
\end{equation}
where $K$ is a constant, the sectional curvature, and the indices
$a,b,c,d$ run only over the $m$D part. Contracting
Eq.~(\ref{mDccRiemann}) yields
\begin{equation}
\label{mDccRicci}
R_{ab} \ = \ K(m-1)\,g_{ab}\, , \qquad R \ = \ Km(m-1)\, ,
\end{equation}
and consequently
\begin{equation}
\label{mDccContrakce}
R_{abcd}^2 \ = \ \frac{2R^2}{m(m-1)}\, , \qquad R_{ab}^2 \ = \ \frac{R^2}{m}\, ,
\end{equation}  
{where} we assumed that $m>1$, since a one-dimensional metric is automatically flat, i.e., has vanishing curvature tensors and scalars. In discussing STs containing an $m$D constant-curvature part, we will use $R$ and $K$ interchangeably, as convenient.

%******
%Note that $d\Sigma_{1,E,K}^2$ can be brought into form $du^2$ (as the only 1D metric possibility). For Lorentzian case, 2 is the minimal dimension for listed form (where $d\Sigma_{2,L,K}^2$ consists only of first two terms). Also note that when suitable (typically when there is only one constant curvature part in metric), we will ``normalize" its $K$ to be 1, 0 or -1 (which is always possible). Constant curvature STs are also maximally symmetric, i.e. they allow maximum number [$m(m+1)/2$] of Killing vector field (KVFs) \cite{STbook, Weinbergbook}. 

Constant-curvature spaces can have either Euclidean or Lorentzian
signature. In the Euclidean case, they correspond to the
$m$D sphere (positive $K$), plane (zero $K$), and hyperbolic space (negative $K$). These three
geometries can be expressed in a unified form, for example, as
\begin{equation}
\label{Ecc}
d\Sigma_{m,E,K}^2 \ = \ \frac{\sum_{a=1}^m du_a^2}
{\left(1+\dfrac{K\sum_{a=1}^m u_a^2}{4}\right)^{2}}\, .
\end{equation}
The Lorentzian case comprises de~Sitter (positive $K$), Minkowski (zero
$K$), and anti-de~Sitter (negative $K$) {geometries}, which can likewise be
unified, e.g., as
\begin{eqnarray}
\label{Lcc}
&&\mbox{\hspace{-10mm}}d\Sigma_{m,L,K}^2 \nonumber \\[2mm] &&\mbox{\hspace{-10mm}}= \ -(1-Kr^2)\,dt^2 \ + \ \frac{dr^2}{1-Kr^2}
\ + \ r^2\, d\Sigma_{m-2,E,1}^2 \, ,
\end{eqnarray}
see also Ref.~\cite{STbook}. Note that $d\Sigma_{1,E,K}^2$ can always be brought to
the form $du^2$, the only possible one-dimensional metric. In the Lorentzian
case, $m=2$ is the minimal dimension to which Eq.~(\ref{Lcc}) applies, with
$d\Sigma_{2,L,K}^2$ then consisting only of its first two terms. We further note that, when convenient --- typically when the metric contains only
a single constant-curvature part --- we normalize $K$ to $+1$, $0$, or $-1$,
which can always be achieved by an appropriate rescaling of the coordinates.
Constant-curvature {metrics} are also maximally symmetric, meaning that they
admit the maximum possible number, $m(m+1)/2$, of Killing vector fields
(KVFs)~\cite{STbook,Weinbergbook}.

Let us now consider what can be said about DP STs composed of the
subspaces described above. First, such {metrics} always admit as many
second-order Killing tensors as there are direct-product parts. These
Killing tensors have nonzero components only within the corresponding
metric {parts}. Another important characteristics concerns the algebraic type of these geometries.
Fortunately, a higher-dimensional generalization of the Petrov classification
is available~\cite{Weylnclass}, and we review it in
Appendix~\ref{APetrov}, where we also use it to characterize the STs
considered here. We find that all of these STs are of either type $D$ or
type $O$. This algebraic classification provides a useful starting point for
interpreting the solutions obtained below.

%%%%%%%%%%%%%%%%%%%%%%%%%%%%%%%%%%%%%%%%%%%
\section{Linear gravity} \label{LG}
%%%%%%%%%%%%%%%%%%%%%%%%%%%%%%%%%%%%%%%%%%%%%

The first theory in which we shall study the models introduced above is
LG, by which we mean a theory whose action is at most linear in the 
curvature scalar terms. In $n$ dimensions, the only curvature scalar of this type is the
Ricci scalar $R$, to which a constant term, corresponding to a cosmological
constant, may also be added. Thus one obtains the $n$-dimensional
Einstein--Hilbert action
\begin{equation}
\label{LGakce}
S_{LG}
\ =\
\frac{1}{\kappa}\int d^n x\,\sqrt{\lvert g\rvert}\,
\left(R-2\Lambda\right)\, ,
\end{equation}
with the ensuing Einstein FEs
\begin{equation}
\label{Einstein}
R_{\mu \nu} \ - \ \frac{1}{2}Rg_{\mu \nu} \ + \ \Lambda g_{\mu \nu} \ \equiv \  G_{\mu \nu} \ + \ \Lambda g_{\mu \nu}  \ = \ 0\, .
\end{equation}
Taking the trace of Eq.~(\ref{Einstein}) gives
\begin{equation}
\label{tracedEinstein}
R
\ =\
\frac{2n}{n-2}\Lambda\, ,
\end{equation}
so that the Ricci scalar is constant.

Let us now consider particular DP STs, beginning with the
$(m_1+m_2)$ structure in which both subspaces have constant curvature.
For $m_2=0$, so that $m_1 \equiv n$, the general solution is simply a
constant-curvature ST with Ricci scalar given by
Eq.~(\ref{tracedEinstein}), corresponding to de Sitter, Minkowski, or
anti-de Sitter ST, according to the sign of $\Lambda$.
For $m_2=1$ ($m_1\equiv m$), the one-dimensional factor has identically
vanishing Ricci tensor and Ricci scalar. The Einstein equations
(\ref{Einstein}), traced over the respective subspaces, give
\begin{equation}
R_{(mD)} \ = \ 2\Lambda\, , \quad R_{(mD)}\ = \ \frac{2\Lambda m }{m-2}\,.
\end{equation}
These two conditions are compatible only for $\Lambda=0$, in which case
$R_{(mD)}=0$. Thus, apart from the trivial Ricci-flat (Minkowski) case, nontrivial DP STs {with} the $(m+1)$ {structure} are not allowed in LG.

For any other values of $m_1$ and $m_2$, Eq.~(\ref{Einstein}) yields two sets of equations, one for each subspace.  If we take the first set, use Eq.~(\ref{mDccRicci}), trace it with the metric of the corresponding subspace, and divide by $m_1$, we obtain
\begin{equation}
\left(\frac{1}{m_1} \ - \ \frac{1}{2}\right)R_{(m_1D)} \  - \ \frac{1}{2}R_{(m_2D)} \ + 
\ \Lambda\ = \ 0\, .
\end{equation}
The equation for the $m_2$D subspace  is obtained analogously by interchanging the indices $1$ and $2$. Solving these two equations gives
\begin{equation}
\label{m1+m2EST}
R_{(m_1D)}\ = \ \frac{2m_1 \Lambda}{n-2}\, , \quad \; R_{(m_2D)} \ = \ \frac{2m_2 \Lambda}{n-2}\, . 
\end{equation}
In the case of the $(2+m)$ model, the procedure works in exactly the same way,
and the results are identical, with $m_1=m$ and $m_2=2$.  This is because, even without assuming the two-dimensional part to have constant curvature, LG does not admit any other solution beyond that implied by
Eq.~(\ref{tracedEinstein}).

For the $(2+\cdots+2)$ structure, the traced FEs for the
$k$th two-dimensional subspace read
\begin{eqnarray}
2\Lambda \ - \ \sum_{i \neq k} R_{(i)}\ = \ 0\, , \quad \; \forall  k \in \  \{1,\ldots,n/2 \,\}.
\end{eqnarray}
Subtracting any two of these equations shows that the corresponding
Ricci scalars must be equal. Hence, all of them are equal and given by
\begin{equation}
\label{2+...+2EST}
R_{(k)} \ = \ \frac{4 \Lambda}{n-2}\, , \quad \; \forall  k \in \{1,\ldots, n/2 \}\,.
\end{equation}
Thus, LG again admits only solutions in which each 2D subspace has constant curvature.

In particular, in 4D, the allowed geometries are the constant-curvature $(4+0)$ STs with $R=4\Lambda$, no $(3+1)$ STs except Minkowski ST, and the $(2+2)$ STs with $R_{(1)}=R_{(2)}=2\Lambda$. In 6D, the allowed geometries are the constant-curvature $(6+0)$ STs with $R=3\Lambda$, no $(5+1)$ STs except Minkowski ST, the $(4+2)$ STs with $R_{(4D)}=2\Lambda=2R_{(2D)}$, the $(3+3)$ STs with $R_{(1)}=R_{(2)}=3\Lambda/2$, and the $(2+2+2)$ STs with $R_{(1)}=R_{(2)}=R_{(3)}=\Lambda$.

%%%%%%%%%%%%%%%%%%%%%%%%%%%%%%%%%%%
\section{Quadratic gravity}
\label{QG}
%%%%%%%%%%%%%%%%%%%%%%%%%%%%%%%%%%%%%%

We now consider a purely quadratic-curvature gravity action in $n$
dimensions of the form
\begin{widetext}
\begin{equation}
\label{QGakce}
S_{QG}
\ =\
\int d^n x\,\sqrt{\lvert g\rvert}\,
\left[
\alpha R^2
\ + \ \beta R_{\mu\nu}^2
\ + \ \gamma\left(
R_{\mu\nu\lambda\sigma}^2
\ - \ 4R_{\mu\nu}^2
\ + \ R^2
\right)
\right]\, ,
\end{equation}
where $\alpha$, $\beta$, and $\gamma$ are arbitrary coefficients.  As noted
above, we exclude the linear and constant terms that would be present in the
most general quadratic-curvature action and restrict our analysis to purely
quadratic terms. The term proportional to $\gamma$ is the Gauss--Bonnet term
(GBT), which is topological in 4D and therefore does not
contribute to the FEs there. The corresponding FEs
are of the form~\cite{QG1,QG2}
\begin{eqnarray}
&&2\alpha R\left(R_{\mu\nu}-\frac{1}{4}R g_{\mu\nu}\right)
\ + \ (2\alpha+\beta)
\left(g_{\mu\nu}\square-\nabla_\mu\nabla_\nu\right)R
\ + \ \beta\square
\left(R_{\mu\nu}-\frac{1}{2}R g_{\mu\nu}\right) \ + \ 2\beta
\left(R_{\mu\lambda\nu\sigma}
-\frac{1}{4}g_{\mu\nu}R_{\lambda\sigma}\right)R^{\lambda\sigma}
\nonumber \\[2mm]
&&
+ \ 2\gamma
\left[
R R_{\mu\nu}
-2R_{\mu\lambda\nu\sigma}R^{\lambda\sigma}
+R_{\mu\lambda\sigma\omega}R_{\nu}^{\ \lambda\sigma\omega}
-2R_{\mu\lambda}R_{\nu}^{\ \lambda}
-\frac{1}{4}g_{\mu\nu}
\left(
R_{\lambda\sigma\omega\vartheta}^2
-4R_{\lambda\sigma}^2
+R^2
\right)
\right]
\ = \ 0\, .
\label{QGfeq}
\end{eqnarray}
Now let us consider some important special cases that will be needed below. 

If the metric has a $(2+\cdots)$ DP structure, the FEs~(\ref{QGfeq}) for the two-dimensional part, upon using
Eqs.~(\ref{2DRiccit})-(\ref{2Dkvad}), reduce to
\begin{eqnarray}
&&
(2\alpha+\beta)
\left[
\left(
\frac{1}{4}R_{(2D)}^2
+\square_{(2D)}R_{(2D)}
+\square_{(\cdots)}R_{(\cdots)}
\right)g_{ij}
\ - \ \nabla_i\nabla_j R_{(2D)}
\right]
\nonumber \\[2mm]
&&
-\ \frac{1}{2}
\left[
\alpha R_{(\cdots)}^2
\ + \ \beta
\left(
R_{\alpha\beta(\cdots)}^2
+\square_{(\cdots)}R_{(\cdots)}
\right)
\ + \ \gamma
\left(
R_{\alpha\beta\gamma\delta(\cdots)}^2
-4R_{\alpha\beta(\cdots)}^2
+R_{(\cdots)}^2
\right)
\right]g_{ij}
\ =\ 0\, ,
\label{QG2D}
\end{eqnarray}
\end{widetext}
where $(\cdots)$ denotes all parts other than the 2D part under discussion, and Greek indices run only over those parts. Note that all terms in Eq.~(\ref{QG2D}), except for the last term in the first line, are proportional to the 2D metric. Therefore, when $2\alpha+\beta\neq0$, the procedure used in Refs.~\cite{previous,Schmidt} can be adapted to the present case. 
In particular, the 2D part admits at least one KVF, so that its metric can be brought to the form
\begin{eqnarray}
ds^{2}_{(2D)} \ = \ \sigma_{(2D)} B(\vartheta)\ \! d\chi^{2}\ + \ B^{-1}(\vartheta)\ \! d\vartheta^{2}\,,  
\label{2Dcanonical}
\end{eqnarray}
with { (due to the fact the same quantity as in~\cite{previous,Schmidt} is subjected to double covariant derivative)}
\begin{eqnarray}
&&B(\vartheta) \ = \ b_0 \ + \ b_1 \vartheta \ + \ b_2\vartheta^2 \ + \ b_3\vartheta^3\, ,\nonumber \\[2mm] 
&&R_{(2D)} \ = \ -B''(\vartheta)\, .
\label{30.kk}
\end{eqnarray}
Here, $\sigma_{(2D)}$ is a signature of the determinant of the 2D metric part $g_{(2D)}$. Explicitly, we have
\begin{eqnarray}
&&\square_{(2D)}R_{(2D)} \ = \ -6b_3\big(3b_3\vartheta^2 \ + \ 2b_2\vartheta \ + \ b_1\big)\, ,\nonumber \\[2mm] 
&& R_{(2D)}\ = \ -\big(6b_3\vartheta \ + \ 2b_2\big)\, .
\label{RicciS2Dcan}
\end{eqnarray}
Note, in particular, that the scalar curvature is independent of $\sigma_{(2D)}$.
Using this result together with the relevant Christoffel symbols
\begin{eqnarray}
&&\Gamma^\vartheta_{\chi\vartheta}
\ = \ 0\,,
\qquad
\Gamma^\vartheta_{\vartheta\vartheta}
\ = \ -\frac{B'(\vartheta)}{2B(\vartheta)}\,,
\nonumber \\[2mm]
&&\Gamma^\vartheta_{\chi\chi}
\ = \ -\frac{1}{2}B'(\vartheta)B(\vartheta)\,,
\end{eqnarray}
and the double covariant derivative 
\begin{eqnarray}
\nabla_i\nabla_j R_{(2D)}
\ = \
\partial_i\partial_j R_{(2D)}
\ - \ \Gamma^k_{ij}\partial_k R_{(2D)}\, .
\end{eqnarray}
allows us to reduce the first line of Eq.~(\ref{QG2D}) to
\begin{eqnarray}
(2\alpha+\beta)
\left[
\square_{(\cdots)}R_{(\cdots)}
\ + \ b_2^2 \ - \ 3b_1b_3
\right]g_{ij}\,.
\end{eqnarray}
For the Lorentzian 2D metric, we will generally identify
$\vartheta\equiv r$ and $\chi\equiv t$, in accordance with
Eq.~(\ref{Lcc}), whereas for the Euclidean metric we make identification
$\vartheta\equiv x$ and $\chi\equiv y$.

For the case $2\alpha+\beta=0$, no quantities associated with the
2D part appear in Eq.~(\ref{QG2D}). Thus, no conclusion about
the geometry of the 2D part can be drawn from this equation alone; the
remaining FEs for the DP ST must also be examined.

If the metric has an $(m+\cdots)$ DP structure, with the $m$D part
having constant curvature, the FEs~(\ref{QGfeq}) for this part
take the form obtained by using Eqs.~(\ref{mDccRiemann})-(\ref{mDccContrakce}):
\begin{widetext}  
\begin{eqnarray}
&&\mbox{\hspace{-9mm}}(2\alpha+\beta) \square_{(\cdots)} R_{(\cdots)}g_{ab} \ + \ \frac{\alpha}{2} \left[ \left(\frac{4}{m}-1\right)R_{(mD)}^2 \ + \ \left( \frac{4}{m}-2\right)R_{(mD)}R_{(\cdots)} \ - \ R_{(\cdots)}^2  \right]g_{ab} \nonumber \\[2mm]
&&\mbox{\hspace{-9mm}}+ \ \frac{\beta}{2}\left[ \left( \frac{4}{m^2}-\frac{1}{m}\right)R_{(mD)}^2 \ - \ R_{\alpha \beta(\cdots)}^2  \ - \ \square_{(\cdots)}R_{(\cdots)}\right]g_{ab} \nonumber \\[2mm]
&&\mbox{\hspace{-9mm}}+ \ \frac{\gamma}{2} \left[ \frac{(4-m)(m-3)(m-2)}{m^2(m-1)}R_{(mD)}^2 \ + \ \left(\frac{4}{m} -2\right)R_{(mD)}R_{(...)} \ - \ \left(R^2_{\alpha \beta \gamma \delta (\cdots)}-4R^2_{\alpha \beta (\cdots)}+R_{(\cdots)}^2 \right) \right]g_{ab} \ = \ 0\, ,~~~
\label{QGmDcc}
\end{eqnarray}
i.e. there is in fact only one independent equation coming from $m$D part, which can be obtained by tracing~(\ref{QGmDcc}) with its metric.

If the metric is of type $(1+\cdots)$, the FE (\ref{QGfeq}) for the 1D part reads
\begin{eqnarray}
\label{QG1D}
\left\{ \left(2\alpha+\beta\right)\square_{(\cdots)}R_{(\cdots)}  \ \! - \ \! \frac{1}{2}\left[\alpha R_{(\cdots)}^2+\beta \left(R_{\alpha \beta(\cdots)}^2+\square_{(\cdots)}R_{(\cdots)}\right)+\gamma\left(R^2_{\alpha \beta \gamma \delta (\cdots)}-4R^2_{\alpha \beta (\cdots)}+R_{(\cdots)}^2 \right)\right] \right\}g_{ii} \ = \ 0\, .~~~
\end{eqnarray}
\end{widetext}
In passing, we note again that the 1D part of the DP metric has identically vanishing curvature. In the present parametrization, we  have $g_{ii}=\pm 1$.

%%%%%%%%%%%%%%%%%%%%%%%%%%%%%%%%%%%%%%%%%%%%%%%%%%%%%%%%%%%%%
\subsection{$(m_1+m_2)$ case for constant curvature subspaces}
\label{Sect.m1_1+m_2}
%%%%%%%%%%%%%%%%%%%%%%%%%%%%%%%%%%%%%%%%%%%%%%%%%%%%%%%%%%%%%

In this case, using the constant-curvature $m$D identities (\ref{mDccRiemann})-(\ref{mDccContrakce}), the FEs~(\ref{QGmDcc}) for the $m_1$D part --- after taking the trace and multiplying by $2/m_1$ ---  become
\begin{widetext}
\begin{eqnarray}
&&\mbox{\hspace{-11mm}}\alpha\left[ \left(\frac{4}{m_1}-1\right)R_{(m_1D)}^2+\left( \frac{4}{m_1}-2\right)R_{(m_1D)}R_{(m_2D)}-R_{(m_2D)}^2  \right] \ + \ \beta\left[ \left( \frac{4}{m_1^2}-\frac{1}{m_1}\right)R_{(m_1D)}^2-\frac{1}{m_2}R_{(m_2D)}^2\right]\nonumber \\[2mm]
&&\mbox{\hspace{-11mm}}+ \ \! \gamma \left[ \frac{(4-m_1)(m_1-3)(m_1-2)}{m_1^2(m_1-1)}R_{(m_1D)}^2+\left(\frac{4}{m_1} -2\right)R_{(m_1D)}R_{(m_2D)}-\left( \frac{2}{m_2(m_2-1)}-\frac{4}{m_2}+1\right)R_{(m_2D)}^2 \right]  =  0\, .~
\label{m1+m2cc}
\end{eqnarray}
\end{widetext}
The FEs for the $m_2$D part are obtained by interchanging $m_1$ and $m_2$. (Here we assume $m_1,m_2\neq 0,1$.) These are algebraic equations that can be solved for $\alpha$, $\beta$, $\gamma$, and $R_{(m_1D)}$, $R_{(m_2D)}$ [or, equivalently, for the corresponding $K$s defined in (\ref{mDccRicci})] once $m_1$ and $m_2$ are fixed. If $m_1$ and $m_2$ are left arbitrary, however, the algebraic degree of the equations and the presence of mixed terms make the resulting general solutions rather cumbersome, and we therefore do not display them. Instead, we focus in the following on several special cases. 

%\textbf{[Dalo by se říct víc, např. vyjádření kvadrátu jednoho R z jedné rce a dosazení do druhé, ale i tak mnoho možností/případů (ne/zápornost různých kořenů apod.) Asi nechat takto?]}

From Sec.~\ref{LG}, we already know the form of LG STs with this structure. We can therefore ask which QG theories admit these LG STs as a subset of their solutions within the DP structure considered here.
To address this question, we substitute Eq.~(\ref{m1+m2EST}) into Eq.~(\ref{m1+m2cc}) and its $m_2$ counterpart. Requiring both expressions to vanish yields a class of QG theories that admit the corresponding LG STs as solutions. This condition leads to three distinct cases:
\begin{eqnarray}
&&\mbox{\hspace{-6mm}}1) \;\;\beta=\frac{-2(\alpha+\gamma)m_1^2+2(\alpha+3\gamma)m_1-6\gamma}{m_1-1}\,, \nonumber \\[2mm] &&\mbox{\hspace{10mm}}{\mbox{with}~~~}m_2=m_1\, ,\nonumber \\[2mm] 
&&\mbox{\hspace{-6mm}}2) \; \;\beta\ = \ -(m_1+m_2) \alpha\, , \;\; \gamma\ = \ 0\, , \nonumber \\[2mm] &&\mbox{\hspace{-6mm}}3) \;\; m_1\ = \ m_2 \ = \ 2\, ,
\label{m1+m2merging}
\end{eqnarray}
where only the fixed constants are displayed, while the remaining constants are arbitrary. In fact, these conditions are sufficiently restrictive to allow all solutions within the class of models under consideration to be determined and expressed in a simple form for each of these QG theories. In particular, Case 3) admits the following general solutions:
\begin{equation}
\label{explLGoverlapA}
R_{(m_1D)} \ = \ \pm R_{(m_2D)}
\qquad \text{or} \qquad
\beta \ = \ -2\alpha\,,
\end{equation}
with both $R_{(m_1D)}$ and $R_{(m_2D)}$ arbitrary. Case 2) admits
\begin{equation}
R_{(m_1D)} \ = \ \frac{m_1 R_{(m_2D)}}{m_2}\, ,
\end{equation}
and case 1) admits
\begin{equation}
R_{(m_1D)} \ = \ R_{(m_2D)}\, ,
\end{equation}
or
\begin{equation}
m_2\ = \ 2,  \;\;\;\; R_{(m_1D)} \ = \ -R_{(m_2D)}\, ,
\end{equation}
or
\begin{equation}
\label{explLGoverlapB}
m_2 \ = \ 2\, ,
 \;\;\;\;
\gamma \ = \ -\alpha\, ,
\end{equation} 
(where  both  $R_{(m_1D)}$ and  $R_{(m_2D)}$ are arbitrary).

There are, of course, overlaps between each pair of these three classes of theories, as well as an overlap of all three cases. The latter occurs, for example when
\begin{equation}
m_1\ = \ m_2 \ = \ 2\, ,
\qquad
\beta\ = \ -4\alpha\, ,
\qquad
\gamma \ = \ 0\, .
\end{equation}
It is worth noting that case 3) implies that, in the $2+2$-dimensional case, every QG theory considered here contains the corresponding LG solutions as a subset of its solutions, within the class of DP models under consideration. We also note that, for $m_1=m_2$, the solutions satisfying
\begin{equation}
R_{(m_2D)} \ = \ -R_{(m_1D)}\, , 
\end{equation}
are conformally flat, as shown in Appendix~\ref{APetrov}.

We now separately consider the case $m_2=0$, i.e., an $(m_1\equiv n)$D constant-curvature ST. Such STs are automatically conformally flat. In this case, all terms with $(\cdots)$ are absent in Eq.~(\ref{QGmDcc}). Taking its trace and multiplying the resulting equation by $2$, we can recast~(\ref{QGmDcc}) in the simple form
\begin{eqnarray}
\label{nDcc}
&&\mbox{\hspace{-10mm}}\frac{R_{(m_1D)}^2(4-m_1)
\left[(\alpha+\gamma)m_1^2 \ + \ (\beta-\alpha-5\gamma)m_1\right]}
{m_1(m_1-1)}
\nonumber\\[2mm]
&&\mbox{\hspace{5mm}}+ \ \frac{R_{(m_1D)}^2(4-m_1)
\left(-\beta+6\gamma\right)}
{m_1(m_1-1)}
\ = \ 0\, .
\end{eqnarray}
The general solution can then be classified into three cases. {First},
\begin{equation}
R_{(m_1D)} \ = \ 0\, ,
\end{equation}
which corresponds to Minkowski ST and imposes no further conditions on the QG parameters. {Second},
\begin{equation}
m_1\ = \ 4\, ,
\end{equation}
which likewise imposes no further conditions. Thus, in  4D, all constant-curvature STs are solutions of any QG theory. 
{Third}, for $R_{(m_1D)}\neq0$ and $m_1\neq4$, the remaining factor must vanish, yielding the condition
\begin{equation}
\label{QGccsol}
\alpha \ = \
\frac{-\gamma m_1^2 \ + \ (5\gamma-\beta)m_1 \ + \ \beta \ - \ 6\gamma}
{m_1(m_1-1)}\, .
\end{equation}\\
Let us also consider the case $m_2=1$, which is also automatically conformally flat, as we show in Appendix~\ref{APetrov}. In this case, we again obtain Eq.~(\ref{nDcc}), but we also have Eq.~(\ref{QG1D}). After taking its trace and multiplying the resulting equation by $-2$, the latter reduces to
\begin{eqnarray}
\label{n-1+1Dcc}
&&\mbox{\hspace{-15mm}}\frac{R_{(m_1D)}^2
\left[(\alpha+\gamma)m_1^2 \ + \ (\beta-\alpha-5\gamma)m_1\right]}
{m_1(m_1-1)}\nonumber \\[2mm]
&&\mbox{\hspace{10mm}}+ \ 
\frac{R_{(m_1D)}^2(-\beta+6\gamma)}
{m_1(m_1-1)}
\ = \ 0\, .
\end{eqnarray}
This equation is identical to Eq.~(\ref{nDcc}) except for the absence of the factor $(4-m_1)$ in the numerator. Consequently, $m_1=4$ is no longer a solution without imposing an additional condition, whereas the other two possibilities identified in the preceding case remain valid.

For $n$D constant-curvature STs, LG solutions are automatically solutions of any theory that admits solutions of this type (and, in particular, for all theories in 4D). For the $(1+m_1)$ type,  LG solutions are only trivial and therefore constitute an obvious subset of the corresponding QG solution set.

%%%%%%%%%%%%%%%%%%%%%%%%%%%%%%%%%%%%%%%%%%%%%%%%%%%%%%%%%%%%%
\subsection{$(2+m)$ case with a constant-curvature $m$D space}
\label{2+mSS}
%%%%%%%%%%%%%%%%%%%%%%%%%%%%%%%%%%%%%%%%%%%%%%%%%%%%%%%%%%%%%%

Here we combine Eqs.~(\ref{QG2D}) and~(\ref{QGmDcc}). First, let us assume that $2\alpha+\beta\neq0$, so that KVF {is present} and the 2D part of the metric can be written in the canonical form (\ref{2Dcanonical}). We begin with Eq.~(\ref{QGmDcc}), which, after taking its trace and dividing by $m$, becomes
\begin{widetext}  
\begin{eqnarray}
&&-(2\alpha+\beta) 6b_3\Big(3b_3\vartheta^2+2b_2\vartheta+b_1\Big) \ + \  \frac{\alpha}{2} \left[ \left(\frac{4}{m}-1\right)R_{(mD)}^2-\left( \frac{4}{m}-2\right)R_{(mD)}\Big(6b_3\vartheta+2b_2\Big)-\Big(6b_3\vartheta+2b_2\Big)^2  \right]\nonumber \\[2mm]
&&\mbox{\hspace{10mm}}+ \ \frac{\beta}{2}\left[ \left( \frac{4}{m^2
}-\frac{1}{m}\right)R_{(mD)}^2 \ - \ 2\Big(3b_3\vartheta+b_2\Big)^2 \ + \ 6b_3\Big(3b_3\vartheta^2+2b_2\vartheta+b_1\Big)\right]
\nonumber \\[2mm]
&&\mbox{\hspace{10mm}}+ \ \frac{\gamma}{2} \left[ \frac{(4-m)(m-3)(m-2)}{m^2(m-1)}R_{(mD)}^2 \ - \ \left(\frac{4}{m} -2\right)R_{(mD)}\Big(6b_3\vartheta+2b_2\Big)\right]\ = \ 0\, ,
\label{2+mQG}
\end{eqnarray} 
\end{widetext}
where we again assumed that $m\neq0,1$. We can regard this equation as a polynomial in $\vartheta$, which requires the coefficient of each power of $\vartheta$ to vanish separately. In particular, the coefficient of the second-order term is $-18b_3^2(3\alpha+\beta)$. The case $b_3=0$ reduces to the case considered in the previous subsection, so here we assume $b_3\neq0$. It then follows that $\beta=-3\alpha.$ With this condition, the coefficient of the first-order term reduces to 
\begin{eqnarray}
-\frac{6b_3R_{(mD)}(m-2)(\alpha+\gamma)}{m} \ = \ 0\, .
\end{eqnarray}
The zeroth-order term yields a more complicated condition. From Eq.~(\ref{QG2D}), we obtain, after taking the trace and dividing by $2$,
\begin{eqnarray}
&&\mbox{\hspace{-10mm}}-\frac{1}{2}\left[ \alpha+\frac{\beta}{m}+\gamma\left( \frac{2}{m(m-1)}-\frac{4}{m}+1\right)\right]R_{(mD)}^2 \nonumber \\[2mm]
&&\mbox{\hspace{15mm}}+ \ (2\alpha+\beta)\Big(b_2^2-3b_3b_1\Big) \ = \ 0\, .
\label{2+mQG2}
\end{eqnarray}
Taking all conditions together (and excluding also $m=2$, since this case is discussed separately), we either have
\begin{equation}
\label{2+m1st}
\beta \ = \ -3\alpha\, , \quad R_{(mD)} \ = \ 0\, , \quad b_2^2 \ = \ 3b_3b_1\, ,
\end{equation}
in which case the second- and first-order coefficients of~(\ref{2+mQG}) vanish, while both the zeroth-order coefficient of~(\ref{2+mQG}) and the zeroth-order condition from~(\ref{2+mQG2}) reduce to the last condition in~(\ref{2+m1st}), or
\begin{equation}
\label{2+m2nd}
\beta \ = \ -3\alpha\, , \quad \gamma \ = \ -\alpha\, , \quad m \ = \ 3\, , \quad b_2^2 \ = \ 3b_3b_1\, ,~
\end{equation}
where the first two conditions again follow from the second- and first-order coefficients of~(\ref{2+mQG}).
Substituting these conditions into the zeroth-order coefficient of~(\ref{2+mQG}) gives
\begin{eqnarray}
-\frac{\alpha}{2}\Big[(m-3)m(m-1)K^2+2(b_2^2-3b_1b_3)\Big]\, , 
\end{eqnarray}
while the corresponding zeroth-order condition from~(\ref{2+mQG2}) gives 
\begin{eqnarray}
-\frac{\alpha}{2}\Big[(m^3-8m^2+19m-12)K^2-2(b_2^2-3b_1b_3)\Big]\, ,~~
\end{eqnarray}
where~(\ref{mDccRicci}) has been used. Their simultaneous vanishing yields the last two conditions in~(\ref{2+m2nd}). In fact, there are additional solutions, but they either merge with cases considered in other subsections or with cases discussed above. Equation (\ref{2+m2nd}) is the only genuinely distinct solution of this type with $n \geq 4$.

We now turn to the case $2\alpha+\beta=0$, i.e. we set
$\beta\equiv-2\alpha$ in the following considerations. Since the existence of a KVF
is not guaranteed in this case, the metric is not necessarily of
the form~(\ref{2Dcanonical}). Eq.~(\ref{QG2D}) then yields an
algebraic condition, obtained by setting the first line of
(\ref{2+mQG2}) to zero, while Eq.~(\ref{QGmDcc}) provides a
differential equation. The algebraic condition following from
Eq.~(\ref{QG2D}) is
\begin{equation}
\label{boxalgebr}
\frac{(m-2)}{m} \left[ \alpha \ + \ \frac{\gamma(m-3)}{m-1}\right]R_{(mD)}^2 \ = \ 0\, .
\end{equation}
Let us first consider the simplest case, $R_{(mD)}=0$, which is allowed for arbitrary $\alpha$, $\gamma$, and $m$. In this case, Eq.~(\ref{QGmDcc}) reduces to $2\alpha\Box_{(2D)}R_{(2D)}=0$, yielding two possibilities:
\begin{eqnarray}
\label{boxrce}
&&\alpha\ = \ 0\, , \quad R_{(mD)} \ = \ 0\, , \nonumber \\[2mm] &&R_{(2D)} \;\; \textrm{ arbitrary}\, ,
\end{eqnarray}
with no restriction on $\gamma$, or
\begin{equation}
\label{onlyover}
R_{(mD)} \ = \ 0\, , \quad \Box_{(2D)}R_{(2D)} \ = \ 0,
\end{equation}
with no restriction on either $\alpha$ or $\gamma$.

Another possibility is to allow $R_{(mD)}$ to be arbitrary and to solve~(\ref{boxalgebr}) in a different way. For $m=2$ this equation is automatically satisfied, and~(\ref{QGmDcc}) reduces to
\begin{eqnarray}
&&m\ = \ 2\, , \quad R_{(mD)} \;\; \textrm{ arbitrary}\,, \nonumber \\[2mm] &&\Box_{(2D)}R_{(2D)} \ = \ 0\, .
\end{eqnarray}
An alternative option is $\alpha=0$, and hence also $\beta=-2\alpha=0$, in which case both $R_{(mD)}$ and $R_{(2D)}$ can be arbitrary, but this theory is trivial in 4D. The case $m=2$ will merge with a wider class of solutions in the next subsection.

The final possibility is to satisfy~(\ref{boxalgebr}) by requiring  the factor $[\ldots]$ vanishes. Together with (\ref{QGmDcc}), this gives
\begin{equation}
\label{boxrce2a}
\alpha=-\frac{\gamma(m-3)}{m-1}\,, \quad CR_{(2D)} \ + \ \Box_{(2D)}R_{(2D)} \ = \ 0\, ,
\end{equation}
with
\begin{equation}
\label{boxrce2b}
m \ \ \neq 3\,, \quad C\ =\ \frac{2(m-2)R_{(mD)}}{m(m-3)}\, ,
\end{equation}
or
\begin{equation}
\label{boxrce3}
\alpha \ = \ 0\, , \quad m\ = \ 3\, , \quad R_{(mD)}R_{(2D)}\ = \ 0\, ,
\end{equation}
with no restriction on $\gamma$. We were not able to obtain general solutions to the differential equations encountered in these cases. We therefore give the explicit forms of the equations, together with some particular solutions, in Appendix~\ref{Abox}.

From Sec.~\ref{LG}, we know that the only LG solutions within the considered DP structure have both curvature scalars $R_{(2D)}$ and $R_{(mD)}$ constant. Thus, the only overlap between the QG and LG solution sets is the case that coincides with a case considered in the previous subsection. For theories, which have $(2+m)$ type LG solutions as a subset while also admitting solutions with a nonconstant-curvature $m$D part, we obtain them by inserting~(\ref{m1+m2EST}) for this structure into~(\ref{2+mQG}) and~(\ref{2+mQG2}). Apart from the general $m=2$ case, this yields the following theories:
\begin{equation}
\label{QGm+2LGasub1}
\beta \ = \ -(m+2)\alpha\, , \quad \gamma \ = \ 0\, ,
\end{equation}
and
\begin{equation}
\label{QGm+2LGasub2}
\alpha\ = \ -\beta/5\, , \quad \gamma\ = \ 0\, ,  \quad m \ = \ 3 \, .
\end{equation}
As can be seen, these cases have a nontrivial overlap only with (\ref{onlyover}).

%%%%%%%%%%%%%%%%%%%%%%%%%%%%%%%%%
\subsection{$(2+\ldots+2)$ case}
\label{2+...+2SS}
%%%%%%%%%%%%%%%%%%%%%%%%%%%%%%%%%%

In this case, we have $n/2$ sets of FEs of the form~(\ref{QG2D}), one for each 2D part, labeled by $(1),\ldots,(n/2)$. The canonical form~(\ref{2Dcanonical}) applies to each 2D part when $2\alpha+\beta \neq 0$, owing to the presence of KVF. First we, however, consider the case $2\alpha+\beta=0$. In this case, the first line of~(\ref{QG2D}) vanishes, while the remaining equations, after taking their trace and dividing by $2$, acquire the  form
\begin{equation}
\label{Yr}
\alpha \sum_{i \neq 1}\square_{(i)}R_{(i)}-(\alpha+\gamma) \sum_{i < j, \: i,j \neq1} R_{(i)}R_{(j)}\ = \ 0\, ,
\end{equation}
for the $(1)$ part of the metric. The corresponding equations for the other 2D parts are analogous. Subtracting one traced equation from another, say the equation associated with factor $(2)$ from that associated with factor $(1)$, gives
\begin{eqnarray}
\label{Xr}
&&\mbox{\hspace{-5mm}}\alpha\big[\square_{(2)}R_{(2)}
\ - \ \square_{(1)}R_{(1)}\big]
\nonumber \\[2mm]
&&\mbox{\hspace{5mm}}- \ (\alpha+\gamma)\sum_{i \neq1,2}R_{(i)} \big[R_{(2)} \ - \ R_{(1)}\big] \ = \  0\, .~~~~
\end{eqnarray}
We can now distinguish several cases. If $\gamma=-\alpha$, Eq.~(\ref{Xr}) implies $ \square_{(2)}R_{(2)}=\square_{(1)}R_{(1)}$.
Since the same subtraction procedure can be applied to any pair of traced equations, all $\square_{(k)}R_{(k)}$ must therefore be equal to same the constant, say $C$. Substituting this result back into~(\ref{Yr}), with $\gamma=-\alpha$, gives $C=0$. We thus obtain the class of solutions
\begin{eqnarray}
\label{2+...+2g=a}
&&\beta \ = \ -2\alpha\, , \qquad \gamma \ = \ -\alpha\, , \qquad \square_{(k)}R_{(k)} \ = \ 0\, , \nonumber \\[2mm] 
&&\forall k \ \in \  \{1, \ldots n/2 \}\, .
\end{eqnarray}
For the explicit form of these equations and some particular solutions, we again refer the reader to Appendix~\ref{Abox}.

The procedure used in~(\ref{Xr}) can be applied to any pair of indices. If not all of the remaining $R$'s are equal, then any $R$ other than those associated with the chosen pair of indices can be eliminated by using this equation. Consequently, such an $R$ must be constant. We have therefore the following option, in which all $R_{(k)}$'s are constant, with their values determined by the $n/2$ equations arising from the second sum in~(\ref{Yr}), namely
\begin{eqnarray}
\label{2+...+2gna1}
&&\mbox{\hspace{-8mm}}\beta \ = \ -2\alpha\, , \nonumber \\[2mm] 
&&\mbox{\hspace{-8mm}}\sum_{i < j, \: i,j \neq k} R_{(i)}R_{(j)} \ = \ 0\, , \qquad \forall k \ \in \ \{1, ...n/2 \}\, .
\end{eqnarray}
Since the $R_{(k)}$'s are constant in this case, we can use~(\ref{2Dcanonical}) for all 2D parts, with $b_3=0$. Thus, according to~(\ref{RicciS2Dcan}) and~(\ref{mDccRicci}), the constants $R_{(i)}$ can be replaced by $b_{2i} ~(\equiv -K_i)$.

Another possibility is that all $R_{(k)}$'s except one are equal to the same constant. This constant must necessarily vanish, so that
\begin{equation}
\label{2+...+2gna2}
\Box R_{(i)} \  = \ 0\, , \qquad R_{(k)} \ = \ 0\, , \qquad \forall k  \ \neq \ i\, . 
\end{equation}
For $2\alpha+\beta \neq 0$, there are (at least) $n/2$ KVFs. We can therefore treat Eqs.~(\ref{QG2D}) for each 2D part as polynomials in $\vartheta_k$, with $k\in{1,\ldots,n/2}$. If all $b_{3k}=0$ --- where the first index labels the coefficient in the canonical form~(\ref{2Dcanonical}) and the second index identifies the corresponding 2D part --- then these polynomials are necessarily constant. Consequently, from~(\ref{QG2D}), after taking the trace and dividing by $2$, we obtain the solutions defined by conditions
\begin{eqnarray}
&&\mbox{\hspace{-15mm}}(2\alpha+\beta)\big[b_{2k}^2-\sum_{i \neq k}b_{2i}^2\big]\nonumber \\[2mm]
&&- \ 4(\alpha+\gamma)\sum_{i < j: \ \! i, j \neq k} b_{2i}b_{2j}\ = \ 0\, , \nonumber \\[3mm]
&&\mbox{\hspace{-15mm}}b_{3k} \ = \ 0\, ,  \qquad \forall k \ \in \ \{1,...n/2 \}\,.
\label{2+...b3k=0}
\end{eqnarray}
These equations can then be solved explicitly once $n$ is specified.

For the case where $b_{3k} \neq 0$ for at least some $k$s,~(\ref{QG2D}) are polynomials of second order in $\vartheta_i$s. The second-order coefficients in the equation for the $(k)$-th part have the form $-18b_{3i}^2(3\alpha+\beta)$ for a single $\vartheta_i^2$ term and $-36b_{3i}b_{3j}(\alpha+\gamma)$ for a mixed $\vartheta_i\vartheta_j$ term, where in both cases $i,j \neq k$. Therefore, we must have $
\beta = -3\alpha$,  $ \gamma  =  -\alpha$. With these relations, the first-  and the second-order terms in $\vartheta_i$ vanish automatically. What remains are the nontrivial zeroth-order terms, which give the class
\begin{eqnarray}
&&\sum_{i \neq k}b_{2i}^2 \ - \ \sum_{i \neq k} 3b_{1i}b_{3i} \ - \ b_{2k}^2+3b_{1k}b_{3k} \ = \ 0\, , \nonumber \\[2mm]
&&\forall  k \ \in \ \{1,\ldots, n/2 \}\, , \nonumber \\[3mm]
&&
\beta \ = \ -3\alpha\, , \qquad \gamma \ = \ -\alpha\, .
\label{2+...+2cond1}
\end{eqnarray} 
In fact, for $n/2 > 2$, summing these equations over all $k$ and comparing the resulting relation with the original equations shows that
\begin{equation}
\label{inbs}
b_{2k}^2 \ - \ 3b_{1k}b_{3k} \ = \ 0\, 
\end{equation}
for all $k$.

Now, let us discuss the intersection between QG and LG in this metric structure. We first consider theories with $2\alpha+\beta=0$. For these theories, inserting~(\ref{2+...+2EST}) into (\ref{QG2D}) gives
\begin{equation}
\left(\frac{n}{2}-1\right)\left(\frac{n}{2}-2\right)(\alpha+\gamma)R_{(k)} \ = \ 0\, .
\end{equation}
Therefore, except in the 4D case, nontrivial LG solutions belong to the solution subset only for QG theories satisfying $\gamma=-\alpha$.

For theories with $2\alpha+\beta\neq 0$, let us use~(\ref{2+...+2EST}) in the form $b_i b_j \equiv C^2$. Then,~(\ref{2+...b3k=0}) takes the form
\begin{equation}
C^2\left(2-\frac{n}{2}\right)
\left[(2\alpha+\beta)+2(\alpha+\gamma)\left(\frac{n}{2}-1\right)\right] \ = \ 0\, .
\end{equation}
Thus, QG theories satisfying
\begin{equation}
\label{2+...+2LGsubstandard}
(2\alpha+\beta) \ + \ 2(\alpha+\gamma)\left(\frac{n}{2}-1\right) \ = \ 0\, ,
\end{equation}
(or {any} 4D theories) have the corresponding LG solutions as a subset of their solutions. These theories have general solutions satisfying the condition
\begin{equation}
\label{LGinQG2+...+2}
\left(\frac{n}{2}-1\right)
\left(b_{2k}^2-\sum_{i \neq k}b_{2i}^2\right)
+2\sum_{i < j: \ \! i,j \neq k}b_{2i}b_{2j} \ = \ 0\, .
\end{equation}
It can be seen that the theories allowing for non-constant 2D parts, described by (\ref{2+...+2cond1}), do not admit the solutions (\ref{2+...+2EST}) (again, except in the 4D case).

In passing, we note that in this paragraph, as well as elsewhere (where analogous counting is required), we use the standard combinatorial rule for selecting two distinct elements from a given set. In particular, when selecting two distinct elements $b_{2i}$ and $b_{2j}$ from the $n/2-1$ elements with $i,j\neq k$, the number of possible choices is
\begin{equation}
\binom{n/2-1}{2}\, .
\end{equation}
%

%%%%%%%%%%%%%%%%%%%%%%
\subsection{4D case \label{IV.D.aa}}
%%%%%%%%%%%%%%%%%%%%%%

We now apply the results obtained in the previous subsections to the 4D case. This case is of particular interest because of its physical relevance and special mathematical properties. It is also the lowest dimension considered in the present analysis and, in this sense, the simplest one. For pure 4D constant-curvature STs, there are no additional conditions, since all such solutions automatically satisfy the QG equations by virtue of condition~(\ref{nDcc}).

A 4D direct-product metric can be either of the form $(1+3)$ or $(2+2)$. Let us first consider the $(1+3)$ case with a 3D constant-curvature space. Then, the only nontrivial condition is (\ref{n-1+1Dcc}), which, for $m_1=3$, is satisfied either for $R_{(3D)}=0$ (i.e., the Minkowski case, which is identical to the previous case) or for $3\alpha+\beta=0$, with an arbitrary constant-curvature 3D part.

The $(2+2)$ case has to be divided into several possibilities. For $2\alpha+\beta\neq 0$, solutions are possible for theories satisfying $\beta=-3\alpha$, subject to the condition~(\ref{2+...+2cond1}), which takes the form
\begin{equation}
\label{4D2+2full}
b_{21}^2 \ - \ 3b_{11}b_{31} \ - \ b_{22}^2 \ + \ 3b_{12}b_{32} \ = \ 0\, ,
\end{equation}
for the first 2D part. The equation for the second 2D part is equivalent to this one and differs only by an overall minus sign. The same case also allows solutions for which both $b_{3k}=0$, subject to the condition~(\ref{2+...b3k=0}), which in 4D takes the form
\begin{equation}
\label{4D2+2bezb3}
\big(2\alpha+\beta\big)\big(b_{21}^2-b_{22}^2\big) \ = \ 0\, .
\end{equation}
Thus, in this case, there is no additional condition on $\alpha$, $\beta$, or $\gamma$.

Finally, for the case $2\alpha+\beta=0$, we do not have KVFs, and therefore we must return to~(\ref{Yr}). In the $(2+2)$ case, this condition simply requires
\begin{equation}
\label{2+2box}
\square R_{(1)} \ = \ 0 \, ,  \qquad
\square R_{(2)} \ = \ 0 \, .
\end{equation}

As for LG solutions, in the $(3+1)$ case we necessarily get $\Lambda=R_{(3D)}=0$, whereas in the $(2+2)$ case we have $ 2\Lambda \ = \ R_{(1)} \ = \ R_{(2)}$. A pure constant-curvature 4D ST,~(\ref{Lcc}), with $4\Lambda=R_{(4D)}$, is an LG solution. Thus, all QG theories within the class of metric structures considered here admit corresponding LG solutions as a subset of their solutions.

We can also ask whether the QG solutions are conformally related to LG solutions, i.e., to solutions of Einstein's FEs. Since the 4D constant-curvature case is already an Einstein ST, we focus on DPs. In~\cite{previous}, we investigated this relation for the $(2+2)$ case in WCG and found that the resulting STs are either themselves solutions of Einstein's FEs or are conformally Einstein, with the corresponding conformal factors determined explicitly. The same conclusion applies here (except for the $2\alpha+\beta=0$ case), since for the combinations of $\alpha$, $\beta$, and $\gamma$ that allow nontrivial solutions, we obtained the same $(2+2)$ metrics defined by relation~(\ref{4D2+2full}). It is worth emphasizing that these conformally Einstein STs also have a known geometrical interpretation: they provide representatives of the C-metric class~\cite{STbook, KinnersleyWalker1970,PlebanskiDemianski1976,KuhnelRademacher2016}.

For the $(3+1)$ case, there are nontrivial QG solutions for $3\alpha+\beta=0$, allowing an arbitrary constant-curvature 3D part. These are particular cases of $(m+1)$ DP metrics, which, according to Appendix~\ref{APetrov}, are conformally flat. We present an explicit Weyl transformation mapping of these solutions to LG solutions in Appendix~\ref{ACEST}. In fact, it can be checked that these STs are themselves of constant curvature: for a Euclidean 3D part, they correspond to de Sitter (or Minkowski) ST, whereas for a Lorentzian 3D part, they correspond to anti-de Sitter (or Minkowski) ST.

Although in this paper we focus on vacuum solutions, the 4D case is sufficiently simple to allow us to investigate electrovacuum solutions in which the electromagnetic (EM)  field respects the same symmetry, i.e., it also has the DP structure. For $(2+\ldots+2)$ metrics, the results can even be generalized. We relegate the details to Appendix~\ref{AEM}; the main result is that, for a $(2+2)$ structure, constant EM-field components associated with both 2D subspaces are allowed. In contrast, for the $(3+1)$ case and for 4D constant-curvature STs, no EM field respecting the assumed structure is allowed. These results can further be generalized to the general $(2+\ldots)$ ST solutions.

%%%%%%%%%%%%%%%%%%%%%%%%%%
\subsection{6D case}
%%%%%%%%%%%%%%%%%%%%%%%%%%%

For a pure constant-curvature 6D ST to exist, the condition
\begin{equation}
\label{QG6+0/5+1c}
-60K^2(30\alpha \ + \ 5\beta \ + \ 12\gamma) \ = \ 0\, ,  
\end{equation}
which follows from~(\ref{nDcc}), must be satisfied. For the $(5+1)$ metric, with the 5D part of constant curvature, the corresponding equation obtained from~(\ref{n-1+1Dcc}) is effectively the same (up to an overall factor of $-2$). For the $(2+4)$ case, with a 4D constant-curvature part, our former discussion shows that, for $2\alpha+\beta\neq 0$, there is only one possibility, namely~(\ref{2+m1st}). For $2\alpha+\beta=0$, the equations for the 2D part derived in Subsection~\ref{2+mSS} and Appendix~\ref{Abox} must instead be solved. In particular, we seek solutions of~(\ref{boxrce}), (\ref{onlyover}), or~(\ref{boxrce2a})--(\ref{boxrce2b}), all for $m=4$.

The $(3+3)$ case, with both parts of constant curvature, is governed by equations of the form~(\ref{m1+m2cc}) and admits a richer set of solutions, namely:
\begin{eqnarray}
&&1)~~~ \gamma \ = \ - \alpha \ = \ \beta/3\, , \nonumber \\[2mm]
&&2)~~~ K_1 \ = \ K_2\, , \qquad \gamma \ = \ -2\alpha \ - \beta/3\, , \nonumber \\[2mm]
&&3)~~~ K_2 \ = \ -K_1\, , \qquad \gamma \ = \ \beta/3\, , \nonumber \\[2mm]
&&4)~~~ K_1 \ = \ K_2 \ = \ 0\, , \nonumber \\[2mm]
&&5)~~~ K_1 \ = \ 0 \, , \qquad \alpha \ = \ -\beta/3\, .
\label{3+3cases}
\end{eqnarray}\\
Here, only the constants that are explicitly constrained by the respective conditions are displayed; all other constants are arbitrary.

The last possibility we consider is the $2+2+2$ structure. For $2\alpha+\beta\neq 0$, with $3\alpha+\beta=0$ and, in general, $b_{3k}\neq 0$, we obtain the conditions~(\ref{2+...+2cond1}). These can be rearranged into separate conditions for each 2D part,
\begin{equation}
\label{2+2+2a}
3b_{3k}b_{1k} \ - \ b_{2k}^2 \ = \ 0\, , \qquad
\forall \, k \ \in \ \{1,2,3\} \, .
\end{equation}
Alternatively, all $b_{3k}$ can vanish. In this case, the conditions~(\ref{2+...b3k=0}) lead to
\begin{eqnarray}
&&\mbox{\hspace{-3mm}}1)~~~ \alpha \ = \ -\beta/2 \ = \ -\gamma\, , \nonumber \\[2mm]
&&\mbox{\hspace{-3mm}}2)~~~ b_{21} \ = \ b_{22} \ = \ b_{23}\, , \quad
\gamma \ = \ -3\alpha/2 \ - \ \beta/4\, , \nonumber \\[2mm]
&&\mbox{\hspace{-3mm}}3)~~~ b_{21} \ = \ b_{22} \ = \ b_{23} \ = \ 0\, , \nonumber \\[2mm]
&&\mbox{\hspace{-3mm}}4)~~~ b_{21}\ = \ b_{22}\ = \ 0\, , \quad
\beta \ = \ -2\alpha\, , \nonumber \\[2mm]
&&\mbox{\hspace{-3mm}}5)~~~ b_{21} \ = \ b_{22}\, , \quad b_{23}\ = \ 0\, , \quad
\gamma \ = \ -2\alpha\ - \ \beta/2\, , \nonumber \\[2mm]
&&\mbox{\hspace{-3mm}}6)~~~ b_{21} \ = \ -b_{22} \ -\ b_{23}\, , \quad
\gamma\ = \ \beta/2\, .
\label{6Dcc2+2+2}
\end{eqnarray}
In cases 4), 5), and 6), the indices labeling the 2D subspaces can, of course, be permuted freely.

The remaining possibility is $2\alpha+\beta=0$. In this case, the
existence of KVFs is not guaranteed, nor can we assume the canonical
2D metric form~(\ref{2Dcanonical}). We must therefore solve
\begin{equation}
\label{6Ddiff1}
\square_{(k)} R_{(k)} \ = \ 0\, , \: \forall k \ = \ \{1, 2, 3 \}\, , \quad \gamma \ = \ -\alpha\, ,
\end{equation}
according to (\ref{2+...+2g=a}), or
\begin{equation}
\label{6Ddiff2}
\square_{(1)} R_{(1)}\ = \ 0\, , \: R_{(2)}\ = \ R_{(3)}\ = \ 0
\end{equation}
according to (\ref{2+...+2gna2}) and (\ref{2+...+2gna1}) [where
the indices can be freely permuted].

We note in passing that case 1) in~(\ref{6Dcc2+2+2}) is listed
even though it satisfies $2\alpha+\beta=0$. It should be emphasized
that this case represents only a particular subset of solutions
within this class, obtained under the assumption that all $R$'s
are constant.

As for the relation to LG, let us first briefly summarize the results obtained for LG solutions in 6D. A 6D constant-curvature ST is automatically an LG solution, with $R=3\Lambda$ according to~(\ref{tracedEinstein}). However, as discussed at the beginning of this subsection, not all QG theories admit these nontrivial constant-curvature solutions.

The $(5+1)$ structure is not allowed as an LG solution unless the 5D part has zero curvature, in which case the entire ST is Minkowski and the solution reduces to the previous case. It is therefore automatically included in the QG solution set. We also know that this metric is conformally flat and can be related to LG solutions by a Weyl transformation; see Appendix~\ref{ACEST}.

For the $(2+4)$ case, Eq.~(\ref{m1+m2EST}) gives one possibility,
\begin{equation}
\label{E2+4}
R_{(4D)} \ = \ 2\Lambda \ = \ 2R_{(2D)}\, ,
\end{equation}\\
which is not a subset of~(\ref{2+m1st}), except in the trivial
case, nor is it a subset of the solutions satisfying
$2\alpha+\beta=0$. Since both parts have constant curvature in
this case, the overlap between the LG and QG solution sets has
already been investigated above. For the $(2+4)$ structure, this
overlap corresponds only to case 2) of~(\ref{m1+m2merging}).

For the $3+3$ case, (\ref{m1+m2EST}) allows only
\begin{equation}
R_{(1)} \ = \ R_{(2)} \ = \ 3\Lambda/2\, ,
\end{equation}
which is contained in cases 1) and 2) of~(\ref{3+3cases}). The corresponding overlap between LG and QG solutions is therefore given by cases 1) and 2) of~(\ref{m1+m2merging}).

Finally, for the $(2+2+2)$ case, (\ref{2+...+2EST}) allows only
\begin{equation}
R_{(1)} \ = \ R_{(2)} \ = \ R_{(3)} \ = \ \Lambda\, ,
\end{equation}
which is contained in cases 1) and 2) of (\ref{6Dcc2+2+2}). The overlap between the LG and QG solution sets was discussed at the end of Subsection~\ref{2+...+2SS}. For the 6D case, the result is that LG solutions are contained in the QG solution subset for theories satisfying
\begin{equation}
\label{LGcontainedinQG}
(2\alpha+\beta) \ + \ 4(\alpha+\gamma) \ = \ 0\, .
\end{equation}

%%%%%%%%%%%%%%%%%%%%%%%%%%%%%%%%%%%%%%%%%%%%
\subsection{Note on Weyl-squared action} \label{V.F.vg}
%%%%%%%%%%%%%%%%%%%%%%%%%%%%%%%%%%%%%%%%%%%%%

Since the present analysis builds in part on our previous work on $(2+2)$ DP STs in WCG~\cite{previous}, and since WCG is an interesting representative of QG, we now briefly introduce this theory and specialize our results to it.

Four-dimensional WCG is based on an action
quadratic in the Weyl tensor. In generic $n$ dimensions, the Weyl tensor is
defined as~\cite{Petersen:16}
\begin{widetext}
\begin{equation}
\label{Weylten}
W_{\mu \nu \rho \sigma} \ = \ R_{\mu \nu \rho \sigma} \ - \ \frac{1}{n-2}(g_{\mu \rho}R_{\nu \sigma}-g_{\mu \sigma}R_{\nu \rho}-g_{\nu \rho}R_{\mu \sigma} \ + \ g_{\nu \sigma}R_{\mu \rho}) \ + \ \frac{1}{(n-1)(n-2)}R(g_{\mu \rho}g_{\nu \sigma}-g_{\mu \sigma}g_{\nu \rho})\, ,
\end{equation}    
\end{widetext}
and thus represents the trace-free part of
the Riemann tensor.

For $n=4$, the WCG action is given by
\begin{eqnarray}
&&\mbox{\hspace{-4mm}}S_{_{\rm W}}[g] =-\frac{1}{4G_{_{\rm W}}^2} \int dx^4 \sqrt{\vert g \vert} \ \! W^{\mu \nu \lambda \kappa} W_{\mu \nu \lambda \kappa} \nonumber \\[2mm]
&&\mbox{\hspace{-4mm}}= -\frac{1}{2G_{_{\rm W}}^2} \int dx^4 \sqrt{\vert g \vert}  \left(\frac{\rm{GBT}}{2} +R_{\mu \nu}R^{\mu \nu}  -  \tfrac{1}{3}R^2\right) ,~~~~~
\label{WCGakce}
\end{eqnarray}
where $G_{_{\rm W}}$ is the WCG coupling constant. $\mathrm{GBT}$, being topological in 4D, does not contribute to the FEs and can therefore be omitted when deriving them.

We emphasize that, since the combination $\sqrt{|g|}\,W^2$ transforms as $\Omega^{n-4}\sqrt{|g|}\,W^2$ under Weyl transformations, it is conformally invariant \emph{only} in $n=4$ dimensions. When extending WCG to higher dimensions, there exist several possible approaches how this can be achieved. One possibility is to retain the quadratic form of the action~(\ref{WCGakce}), including the GBT, which is no longer topological for $n>4$. This keeps the theory within the QG class, at the price of losing conformal invariance. Another possibility is to modify the action so that conformal invariance is preserved in higher dimensions; this specific route will be analyzed in our follow-up work. Here, we adopt the first approach and treat the Weyl-squared action as a particular example of QG.

Let us now discuss this theory in the context of QG. In $n$ dimensions, the square of the Weyl tensor can be decomposed as
\begin{equation}
W^2 \ = \ R_{\mu \nu \sigma \rho}^2 \ - \ \frac{4R_{\mu \nu}^2}{n-2} \ + \ \frac{2R^{2}}{(n-2)(n-1)}\,  ,  
\end{equation}
so that the corresponding coefficients in the QG action~(\ref{QGakce}) are
\begin{equation}
\label{coeffCG}
\alpha \ = \ \frac{n(3-n)}{(n-2)(n-1)}\, , \quad  \beta \ = \  \frac{4(n-3)}{n-2}\, , \quad \gamma \ = \ 1\, .
\end{equation}
We now examine the solutions with the DP structure discussed previous sections for this particular choice of QG coefficients. For the $(m_1+m_2)$ case, the only solution of the equations~(\ref{m1+m2cc}) is
\begin{equation}
\label{CFagain}
R_2 \ = \ -\frac{R_1m_2(m_2-1)}{m_1(m_1-1)}\, ,
\end{equation}
which corresponds to the conformally flat case. In addition, all $n$D constant-curvature STs, as well as the $(m+1)$ configurations with a constant-curvature $m$D part, are admissible solutions: these geometries are conformally flat, so that $W_{\mu\nu\rho\sigma}=0$ and the FEs are satisfied identically. Equivalently, conditions~(\ref{nDcc}) and~(\ref{n-1+1Dcc}) are always met for the coefficients~(\ref{coeffCG}).

For the $2+m$ case, the relation $\beta=-3\alpha$, which is necessary for the existence of solutions with nonconstant two-dimensional curvature, is not satisfied by the coefficients~(\ref{coeffCG}) for $n\neq4$. Consequently, only solutions with constant 2D curvature are possible, and these reduce to the cases considered in Sec.~\ref{Sect.m1_1+m_2} { and therefore (\ref{CFagain}) with $m_2 \equiv 2$}. The same conclusion applies to the $(2+\ldots+2)$ case: only the solutions described by Eq.~(\ref{2+...b3k=0}) are admissible .

The 4D WCG theory corresponds to $\alpha=-2/3$, $\beta=2$, and $\gamma=1$ (the GBT can be omitted, since it does not contribute to FEs in 4D). As discussed in Sec.~\ref{QG}, there are no restrictions on $(4+0)$ constant-curvature STs; all such solutions are allowed. Likewise, there are no restrictions on the $(3+1)$ case with a 3D constant-curvature factor, since $3\alpha+\beta=0$. The $(2+2)$ case admits the solutions~(\ref{4D2+2full}). This case was already discussed in detail in Ref.~\cite{previous}, including its conformal relations to other WCG solutions and to LG solutions; these results therefore apply also to the corresponding QG solutions~(\ref{4D2+2full}). Such relations are particularly important in conformal gravity, since conformal invariance ensures that every metric Weyl-related to a solution is again a solution.

From the previous discussion it follows that, in 6D (with $\alpha=-9/10$, $\beta=3$, $\gamma=1$), all $(6+0)$ and $(5+1)$ constant-curvature STs are allowed. For $(3+3)$, one has $R_1=-R_2$. In the $(2+4)$ case only constant-curvature 2D parts are allowed, with $R_{(2D)}=-R_{(4D)}/6$, and in the $(2+2+2)$ case no solutions exist except Minkowski ST.

%%%%%%%%%%%%%%%%%%%%%%%%%%%%%%%%%%%%%%%
\subsection{Note on lower dimensions} \label{lowerdim}
%%%%%%%%%%%%%%%%%%%%%%%%%%%%%%%%%%%%%%%

Although throughout the text we have assumed $n\geq4$, let us briefly comment on lower dimensions. In 2D, owing to the identities~(\ref{2DRiemann})--(\ref{2Dkvad}), the Riemann and Ricci tensors are completely determined by the Ricci scalar and the metric tensor.  Since the Einstein tensor in~(\ref{Einstein}) vanishes identically in 2D, LG is trivial there and is usually replaced by other theories, most notably dilaton gravity (such as the Jackiw--Teitelboim model), see, e.g., Ref.~\cite{Teitelboim1983,Jackiw1985,GKV2002}, which are not the subject of this paper. In this case, GBT combination vanishes identically, and the integrand of the QG action reduces to $(\alpha+\beta/2)R^2$, i.e., the theory becomes a particular case of $f(R)$ gravity~\cite{HJS2D}.

In 3D, LG has no propagating degrees of freedom, but it admits nontrivial solutions. Within the models considered here, only the constant-curvature $(3+0)$ and the general $(2+1)$ structures can be studied. In the first case, all constant-curvature STs with $R=6\Lambda$ are allowed. In the latter case, the FEs force $\Lambda=0$ and $R_{(2D)}=0$, so that only Minkowski ST remains, which is contained in the previous case. As for QG, since the Riemann tensor in 3D is determined by the Ricci tensor, the Ricci scalar, and the metric~\cite{book2}, GBT vanishes identically, and only the terms $\alpha R^2$ and $\beta R_{\mu \nu}^2$ remain in the integrand, with $\gamma$ arbitrary. In principle, {this theory is part of our previous} QG analysis, but we do not pursue it here, since neither the Weyl tensor nor the generalized Petrov classification is available in 3D.

As for the Weyl-squared action, used in Sec.~\ref{V.F.vg} as a particular example of QG, the Weyl tensor vanishes identically for $n<4$, so this theory is trivial in these dimensions. 
%In 3D, a conformally invariant theory can nevertheless be constructed from the gravitational Chern--Simons term, whose FEs state the vanishing of the Cotton tensor~\cite{DJT1982,HorneWitten1989}; its solutions are precisely the conformally flat metrics. In 2D, every metric is locally conformally flat, and no nontrivial local conformal gravity exists.

%

%%%%%%%%%%%%%%%%%%%%%%%%%%%%%
\section{Conclusion}
\label{concl}
%%%%%%%%%%%%%%%%%%%%%%%%%%%%%%

DP STs offer a useful middle ground between the full complexity of the higher-dimensional gravitational FEs and the limited scope of highly specific ansatzes. In this work, we exploited this balance to study DP geometries of the $(2+\ldots+2)$, $(2+m)$, and $(m_1+m_2)$ types --- the last two with constant-curvature $m$D parts ---  in LG and QG. The resulting geometries retain sufficient structure to exhibit nontrivial curvature and conformal properties, while their FEs reduce to a form that can be analyzed systematically. We also found that all DP STs considered here are algebraically special, being of generalized Petrov type D or O, which provides a useful geometric characterization of the entire class.

LG provides a particularly simple starting point. For the DP structures considered here, the vacuum LG equations (with a cosmological constant) are restrictive enough to force the 2D parts to have constant curvature. Thus, although a general 2D metric was allowed initially, LG does not admit genuinely nonconstant-curvature solutions within the DP ansatz. The resulting solutions are characterized by a small set of curvature constants fixed by the cosmological constant and the dimensions of the constituent subspaces. This simple structure provides a useful benchmark  for identifying which features of the DP geometries arise specifically from higher-curvature terms.

QG leads to a substantially richer picture. For the classes of metrics considered here, we derived the reduced FEs and used them to determine the conditions on the  coefficients $\alpha$, $\beta$, and $\gamma$ (which specify the particular QG theory) and on the ST dimension under which nontrivial DP solutions exist. We also gave the explicit relations that define these solutions. For constant-curvature subspaces, these relations are algebraic, which makes a systematic classification possible. In this sense, the DP ansatz makes explicit how the three quadratic-curvature {scalars} enter the FEs and how their relative weights depend on the dimension.  When a (generally nonconstant-curvature) 2D part is present, two distinct cases arise. If $2\alpha+\beta\neq0$, the 2D metric is guaranteed to admit  KVF and can therefore be brought to a canonical form, in which the equations determining the solution again become algebraic. The special locus $2\alpha+\beta=0$ in the space of couplings stands apart, because there the existence of KVF can no longer be assumed, and the equations become differential. In this sector we obtained several explicit branches, but a general form of its solutions remains open. 

An important part of our analysis was the relation between QG and LG. Rather than merely testing whether a given DP geometry solves both theories, we determined the conditions under which the space of LG solutions is contained in the corresponding QG solution space, and identified cases in which additional QG branches appear. The 4D theory is particularly special in this respect. There, constant-curvature STs solve both the LG (i.e., Einstein) equations and the pure QG equations without further restrictions, while, for $2\alpha+\beta\neq0$, the  $(2+2)$ sector contains solutions that are either Einstein or conformally Einstein. The latter are also closely related to the C-metric class. We also found that the $(m+1)$ DP geometries with a constant-curvature $m$D factor are conformally flat and admit explicit Weyl transformations relating them to LG solutions [which includes the $(3+1)$ case]. These observations illustrate that, even within the highly constrained DP setting, higher-curvature gravity can enlarge the solution space without necessarily producing geometries unrelated to Einstein gravity.  Besides the 4D case, we also treated the 6D case, which represents a tractable and widely studied higher-dimensional example, for which we presented the solutions explicitly.  In addition to these examples specified by the dimension, we studied a particular choice of couplings, namely the Weyl-squared action, which reduces to WCG for $n=4$ but is not conformally invariant for $n>4$.

The inclusion of an EM field further demonstrates the usefulness of the DP construction. In 4D, the $(2+2)$ geometry permits constant EM field components associated with the 2D parts, while the corresponding $(3+1)$ and $(4+0)$ constant-curvature configurations do not admit a nontrivial EM field respecting the assumed DP structure. The same structural argument extends to the general $(2+\ldots+2)$ class in QG with $2\alpha+\beta\neq0$, where each 2D part can carry its own constant EM field component.

Taken together, these results highlight a useful distinction between the two theories considered here. LG strongly constrains the DP parts and essentially collapses the general two-dimensional sectors to constant curvature. QG admits a much broader family of DP solutions whose existence can, for the most part, be classified algebraically in terms of the couplings and the ST dimension. The Weyl-squared example shows how strongly a specific choice of couplings can restrict this family: for $n>4$, it admits only conformally flat $(m_1+m_2)$ solutions, excludes nonconstant-curvature 2D parts, and, in 6D, leaves only Minkowski ST in the $(2+2+2)$ sector. The DP setting thus provides a particularly transparent way of seeing both the similarities and the differences between these theories. 

Several directions naturally suggest themselves for future work. One is to enlarge the geometric ansatz and investigate DP structures beyond those considered here, particularly combinations involving more general nonconstant-curvature components. Another is to study the mathematical and physical properties of the solutions already obtained in greater detail, including their possible physical interpretation. It would also be worthwhile to pursue the conformal relations between the different solution branches more systematically, especially in dimensions higher than four. Even more generally, extending the present vacuum analysis to non-vacuum DP solutions may reveal further connections between higher-curvature gravity, conformal geometry, and physically relevant higher-dimensional STs. In a follow-up work, we will investigate higher-dimensional conformal gravity. In higher dimensions, conformally invariant actions are no longer unique: several independent invariants exist, and, in general, both the form of a polynomial action and the order of the resulting FEs depend on the dimension~\cite{Lu2013, Anastasiou2021, BoulangerErdmenger2004, BoulangerRovere2025}. To carry out an analysis in the same manner as here, one needs a universal action in which the dimension $n$ enters only as a numerical parameter. Such a theory was recently introduced in Ref.~\cite{Ndimconf}; its action is polynomial only when $n$ is divisible by four, but its FEs remain of fourth order in every dimension.

%%%%%%%%%%%%%%%%%%%%%%%%%%%%%
\begin{acknowledgments}
%%%%%%%%%%%%%%%%%%%%%%%%%%%%%
%
We thank Dr.~Alena Pravdov\'{a} for enlightening discussion on generalized Petrov classification.
P.J. was supported by the Czech Science Foundation Grant (GA\v{C}R), Grant No. 25-18105S. T.L. was supported by the Grant Agency of the Czech Technical University in Prague, grant No. SGS25/163/OHK4/3T/14.
\end{acknowledgments}

\appendix

%%%%%%%%%%%%%%%%%%%%%%%%%%%%%%
\section{
Generalized Petrov classification and its application to the geometries considered}
\label{APetrov}
%%%%%%%%%%%%%%%%%%%%%%%%%%%%%%

For the generalized ($n$-dimensional) Petrov classification, the algebraic structure of the Weyl tensor is characterized in terms of its alignment with preferred null directions~\cite{Weylnclass,Weylnclass2}. This generalizes the 4D situation, where the Petrov classification is based on the existence and multiplicity of principal null directions. 

The basic idea is to determine whether there exists a null direction along which the Weyl tensor satisfies increasingly restrictive alignment conditions. The degree of this alignment is quantified by the \emph{alignment order}, and the different possible alignment orders define the generalized Petrov types. 

To implement this classification, one introduces a frame consisting of \(n\) vectors,
\begin{eqnarray}
\left\{\mathfrak{l},\mathfrak{n},\mathfrak{m}^{(\alpha)}\right\}\, ,
\qquad
\alpha \ = \ 2,\ldots,n-1\, ,
\end{eqnarray}
where $\mathfrak{l}$ and $\mathfrak{n}$ are null vectors and the $n-2$ vectors $\mathfrak{m}^{(\alpha)}$ are spacelike. The frame is chosen to satisfy
\begin{eqnarray}
&&\mathfrak{n}^\mu \mathfrak{n}_\mu \ = \  \mathfrak{l}_\mu \mathfrak{l}^\mu \ = \ \mathfrak{n}^\mu \mathfrak{m}_\mu^{(\alpha)}\ = \ \mathfrak{l}^\mu \mathfrak{m}_\mu^{(\alpha)} \ = \ 0\, ,\nonumber \\[2mm]
&&\mathfrak{n}_\mu \mathfrak{l}^\mu \ = \ 1, \quad \mathfrak{m}_\mu^{(\alpha)}\mathfrak{m}^{\mu \:(\beta)} \ = \ \delta_{\alpha \beta}\, .
\label{n-ady}
\end{eqnarray}
The Weyl tensor components are then expressed with respect to this frame. We denote the vector $\mathfrak{l}$, which is selected as the null direction of highest alignment order, by the index $0$, the second null vector $\mathfrak{n}$ by the index $1$, and the spacelike vectors by Latin indices $i,j,\ldots$. 
This notation should not be confused with the two-dimensional indices introduced in Sect.~\ref{def,id} or with other index conventions used earlier. Thus, for example
\begin{eqnarray}
&&W_{0i0j}
\ \equiv \ 
W_{\mu\nu\rho\sigma} \,
\mathfrak{l}^{\mu} \,
\mathfrak{m}^{(i)\nu} \,
\mathfrak{l}^{\rho} \,
\mathfrak{m}^{(j)\sigma}\, ,\nonumber \\[2mm]
&&W_{01ij}
\ \equiv \ 
W_{\mu\nu\rho\sigma} \,
\mathfrak{l}^{\mu} \,
\mathfrak{n}^{\nu} \,
\mathfrak{m}^{(i)\rho} \,
\mathfrak{m}^{(j)\sigma}\, ,
\end{eqnarray}
denote the projections of the Weyl tensor onto the corresponding frame vectors.

The notion of alignment can be made explicit by considering 
a Lorentz boost in the $\mathfrak l$--$\mathfrak n$ plane, i.e.
\begin{equation}
\mathfrak l\longrightarrow \lambda\mathfrak l\, ,
\qquad
\mathfrak n\longrightarrow \lambda^{-1}\mathfrak n\, ,
\qquad
\mathfrak m^{(i)}\longrightarrow\mathfrak m^{(i)}\, ,
\end{equation}
%
%where $\lambda>0$. Under such a transformation, different Weyl tensor components acquire different powers of $\lambda$, referred to as their boost weights. 
%A null direction is said to be aligned with the Weyl tensor if the components with the highest boost weights vanish in a frame adapted to that direction. The alignment order specifies the successive levels of boost weight whose components vanish.
%
where $\lambda>0$. Under such a transformation, a Weyl tensor component acquires a factor $\lambda^{b}$, where the integer $b$ (the number of indices $0$ minus the number of indices $1$) is called its boost weight; for the Weyl tensor, $b=-2,\ldots,+2$. A null direction $\mathfrak l$ is said to be aligned with the Weyl tensor if the components of the highest boost weight, $b=+2$ (e.g., $W_{0i0j}$), vanish in a frame adapted to that direction. The alignment order is the number of successive levels of boost weight, counted from the highest one downwards, whose components vanish, see Tab.~\ref{tab:boost}.

%
%\begin{widetext}
% Usage: \begin{figure}[t]\centering \input{boost_figure_snippet} \caption{...}\end{figure}
% Needs in the preamble: \usepackage{tikz}, \usepackage{amsmath,amssymb}
% Usage: \begin{figure}[t]\centering \input{boost_figure_snippet} \caption{...}\end{figure}
%
\begin{widetext}
\begin{center}
\begin{table}[H]
\centering
\begin{tikzpicture}[font=\small, x=1cm, y=1cm]
  % ---------------- geometry ----------------
  \def\xb{0.0}      % left edge of the table
  \def\xc{1.55}     % left end of component column
  \def\xt{6.45}     % left end of the first type column
  \def\w{1.25}      % width of a type column
  \def\h{0.95}      % row height
  \pgfmathsetmacro{\xr}{\xt+5*\w}   % right edge of the table

  % ---------------- header ----------------
  \node[anchor=west] at (\xb,1.45) {\textbf{$b$}};
  \node[anchor=west] at (\xc,1.45) {\textbf{Weyl components}};
  \foreach \j/\T in {0/G,1/I,2/II,3/III,4/N}{
    \node at ({\xt+\w*(\j+0.5)},1.45) {\textbf{\large \T}};
    \node at ({\xt+\w*(\j+0.5)},0.82) {\footnotesize order $\j$};
  }
  \node[anchor=east,align=right,font=\footnotesize] at (\xt-0.1,1.45)
        {type $\rightarrow$};
  \node[anchor=east,align=right,font=\footnotesize] at (\xt-0.1,0.82)
        {alignment order $\rightarrow$};
  \draw[thick] (\xb,1.15) -- (\xr,1.15);

  % ---------------- rows ----------------
  \foreach \r/\b/\comp in {
      0/{+2}/{$W_{0i0j}$},
      1/{+1}/{$W_{0ijk},\ W_{010i}$},
      2/{0}/{$W_{0101},\ W_{01ij},\ W_{0i1j},\ W_{ijkl}$},
      3/{-1}/{$W_{1ijk},\ W_{101i}$},
      4/{-2}/{$W_{1i1j}$}}{
    \pgfmathsetmacro{\y}{-\h*\r}
    \node[anchor=west] at (\xb,\y) {$b=\b$};
    \node[anchor=west] at (\xc,\y) {\comp};
    \foreach \j in {0,...,4}{
      \ifnum\r<\j
        \fill[gray!35] ({\xt+\w*\j+0.04},{\y-0.4}) rectangle ({\xt+\w*(\j+1)-0.04},{\y+0.4});
        \node at ({\xt+\w*(\j+0.5)},\y) {$0$};
      \else
        \fill[blue!14] ({\xt+\w*\j+0.04},{\y-0.4}) rectangle ({\xt+\w*(\j+1)-0.04},{\y+0.4});
        \node at ({\xt+\w*(\j+0.5)},\y) {$\neq0$};
      \fi
    }
  }
  % light separators between rows
  \foreach \r in {0,...,3}{
    \draw[black!25] (\xb,{-\h*\r-\h/2}) -- (\xr,{-\h*\r-\h/2});
  }
  \draw[thick] (\xb,{-4*\h-0.5}) -- (\xr,{-4*\h-0.5});

  % ---------------- sign of b, on the right ----------------
  \draw[black!55] ({\xr+0.15},{0.4}) -- ({\xr+0.15},{-\h-0.4});
  \node[anchor=west,font=\footnotesize,text=black!70] at ({\xr+0.2},{-0.5*\h}) {$b>0$};
  \draw[black!55] ({\xr+0.15},{-2*\h+0.4}) -- ({\xr+0.15},{-2*\h-0.4});
  \node[anchor=west,font=\footnotesize,text=black!70] at ({\xr+0.2},{-2*\h}) {$b=0$};
  \draw[black!55] ({\xr+0.15},{-3*\h+0.4}) -- ({\xr+0.15},{-4*\h-0.4});
  \node[anchor=west,font=\footnotesize,text=black!70] at ({\xr+0.2},{-3.5*\h}) {$b<0$};
\end{tikzpicture} 
\vspace{4mm}
\caption{\small{
Boost-weight structure of the Weyl tensor and the \emph{principal} algebraic classification (types G, I, II, III, N), in which only the alignment of the null vector $\mathfrak l$ is considered. A boost $\mathfrak l\to\lambda\mathfrak l,\ \mathfrak n\to\lambda^{-1}\mathfrak n$ multiplies a component by $\lambda^{b}$, with $b=\#(\text{indices }0)-\#(\text{indices }1)$. Grey cells (0) vanish in a frame adapted to $\mathfrak l$; blue cells may be nonzero. The alignment order is the number of vanishing levels, counted from $b=+2$ downwards. 
}}
\label{tab:boost}
\end{table}
\end{center}
\end{widetext}

The primary classification is therefore determined by the highest alignment order attainable for a null direction. The generic case, for which no aligned null direction exists, is denoted by type G. Type I is obtained when there exists a null direction for which
\begin{eqnarray}
W_{0i0j} \ = \ 0\, .
\end{eqnarray}
For type II, the alignment is stronger and, in addition,
\begin{eqnarray}
W_{0ijk} \ = \ 0\, .
\end{eqnarray}
Note that, since the Weyl tensor is trace-free, type~II automatically also includes the case like $W_{010j}=0$. Indeed, in the null frame~(\ref{n-ady}), $0=g^{ab}W_{a0cj}=W_{100j}+\sum_iW_{i0ij}$, so that, by the antisymmetry in the first pair, $W_{010j}=-\sum_iW_{0iij}$. This is a trace of $W_{0ijk}$, and so $W_{0ijk}=0$ implies $W_{010j}=0$. A similar remark is valid also for other types.

For type III, one further requires
\begin{eqnarray}
W_{ijkl} \ = \ W_{01ij} \ = \ 0\, ,
\end{eqnarray}
while for type N one additionally has
\begin{eqnarray}
W_{1ijk} \ = \ 0\, .
\end{eqnarray}
Accordingly, the alignment order of the maximally aligned null direction is 1, 2, 3, and 4 for types I, II, III, and N, respectively, see Tab.~\ref{tab:boost}. Finally, type O is distinguished from these cases by the complete absence of Weyl curvature,
\begin{eqnarray}
W_{\mu\nu\rho\sigma} \ = \ 0\, ,
\end{eqnarray}
and therefore corresponds to conformally flat ST.

The primary classification can be further refined by examining the alignment of the Weyl tensor with the second null direction, $\mathfrak{n}$. This leads to the secondary classification, in which a type is characterized by the alignment orders associated with the two null directions. The resulting classes are denoted by pairs $(p,q)$, where $p$ and $q$ are the alignment orders of the $0$-th and $1$-st null directions, respectively. The cases relevant here include
\begin{eqnarray}
(1,1)\, ,\qquad (2,1)\, ,\qquad (3,1)\, ,\qquad (2,2)\, .
\end{eqnarray}
The class $(2,2)$ represents the special case in which both null directions have alignment order two. It is a special subclass of \((2,1)\) and is conventionally referred to as \emph{type} D, cf. Ref.~\cite{Weylnclass2}.
For more details, see, Tab.~\ref{tab:boost2}. 

Having reviewed the relevant aspects of the generalized Petrov classification, we now apply it to the cases discusses in this paper.

\begin{widetext}
\begin{center}
\begin{table}[H]
\centering
\begin{tikzpicture}[font=\small, x=1cm, y=1cm]
  % ---------------- geometry ----------------
  \def\xb{0.0}      % left edge of the table
  \def\xc{1.55}     % left end of component column
  \def\xt{6.45}     % left end of the first type column
  \def\w{1.25}      % width of a type column
  \def\h{0.95}      % row height
  \pgfmathsetmacro{\xr}{\xt+4*\w}   % right edge of the table

  % ---------------- header ----------------
  \node[anchor=west] at (\xb,1.45) {\textbf{$b$}};
  \node[anchor=west] at (\xc,1.45) {\textbf{Weyl components}};
  \foreach \j/\T/\o in {0/{I}/{(1,1)},1/{II}/{(2,1)},2/{III}/{(3,1)},3/{D}/{(2,2)}}{
    \ifnum\j<3
      \node at ({\xt+\w*(\j+0.5)},1.45) {\textbf{\large \T}$_{\boldsymbol{i}}$};
    \else
      \node at ({\xt+\w*(\j+0.5)},1.45) {\textbf{\large \T}};
    \fi
    \node at ({\xt+\w*(\j+0.5)},0.82) {\footnotesize \o};
  }
  \node[anchor=east,font=\footnotesize] at (\xt-0.1,1.45) {type $\rightarrow$};
  \node[anchor=east,font=\footnotesize] at (\xt-0.1,0.82) {alignment type $(k,l)$ $\rightarrow$};
  \draw[thick] (\xb,1.15) -- (\xr,1.15);

  % ---------------- row labels and components ----------------
  \foreach \r/\b/\comp in {
      0/{+2}/{$W_{0i0j}$},
      1/{+1}/{$W_{0ijk},\ W_{010i}$},
      2/{0}/{$W_{0101},\ W_{01ij},\ W_{0i1j},\ W_{ijkl}$},
      3/{-1}/{$W_{1ijk},\ W_{101i}$},
      4/{-2}/{$W_{1i1j}$}}{
    \pgfmathsetmacro{\y}{-\h*\r}
    \node[anchor=west] at (\xb,\y) {$b=\b$};
    \node[anchor=west] at (\xc,\y) {\comp};
  }

  % ---------------- cells: first all "may be nonzero", then the vanishing levels in grey ----------------
  \foreach \j in {0,...,3}{ \foreach \r in {0,...,4}{
    \pgfmathsetmacro{\y}{-\h*\r}
    \fill[blue!14] ({\xt+\w*\j+0.04},{\y-0.4}) rectangle ({\xt+\w*(\j+1)-0.04},{\y+0.4});
    \node at ({\xt+\w*(\j+0.5)},\y) {$\neq0$};
  }}
  % I_i: b=+2,-2 ;  II_i: b=+2,+1,-2 ;  III_i: b=+2,+1,0,-2 ;  D: b=+2,+1,-1,-2
  \foreach \j/\van in {0/{0,4}, 1/{0,1,4}, 2/{0,1,2,4}, 3/{0,1,3,4}}{
    \foreach \r in \van{
      \pgfmathsetmacro{\y}{-\h*\r}
      \fill[gray!35] ({\xt+\w*\j+0.04},{\y-0.4}) rectangle ({\xt+\w*(\j+1)-0.04},{\y+0.4});
      \node at ({\xt+\w*(\j+0.5)},\y) {$0$};
    }}

  % separators
  \foreach \r in {0,...,3}{
    \draw[black!25] (\xb,{-\h*\r-\h/2}) -- (\xr,{-\h*\r-\h/2});
  }
  \draw[thick] (\xb,{-4*\h-0.5}) -- (\xr,{-4*\h-0.5});

  % ---------------- sign of b, on the right ----------------
  \draw[black!55] ({\xr+0.15},{0.4}) -- ({\xr+0.15},{-\h-0.4});
  \node[anchor=west,font=\footnotesize,text=black!70] at ({\xr+0.2},{-0.5*\h}) {$b>0$};
  \draw[black!55] ({\xr+0.15},{-2*\h+0.4}) -- ({\xr+0.15},{-2*\h-0.4});
  \node[anchor=west,font=\footnotesize,text=black!70] at ({\xr+0.2},{-2*\h}) {$b=0$};
  \draw[black!55] ({\xr+0.15},{-3*\h+0.4}) -- ({\xr+0.15},{-4*\h-0.4});
  \node[anchor=west,font=\footnotesize,text=black!70] at ({\xr+0.2},{-3.5*\h}) {$b<0$};
\end{tikzpicture}
\vspace{4mm}
\caption{Secondary classification of the Weyl tensor, complementing Tab.~\ref{tab:boost}. In addition to $\mathfrak l$, the second null vector $\mathfrak n$ is aligned, so that, besides the conditions of the corresponding principal type, one also has $W_{1i1j}=0$ (for type D, all $b=-1$ components vanish as well). Each type is characterized by the pair $(k,l)$ of alignment orders of $\mathfrak l$ and $\mathfrak n$: $\mathrm{I}_i=(1,1)$, $\mathrm{II}_i=(2,1)$, $\mathrm{III}_i=(3,1)$, $\mathrm D=(2,2)$. A boost $\mathfrak l\to\lambda\mathfrak l,\ \mathfrak n\to\lambda^{-1}\mathfrak n$ multiplies a component by $\lambda^{b}$, with $b=\#(\text{indices }0)-\#(\text{indices }1)$. Grey cells (0) vanish in the adapted frame; blue cells may be nonzero. In type D only the $b=0$ components survive.}
\label{tab:boost2}
\end{table}
\end{center}
\end{widetext}
%

%%%%%%%%%%%%%%%%%%%%%%%%%%%%%%%
\subsection{$(2+\ldots+2)$ metric structure}
%%%%%%%%%%%%%%%%%%%%%%%%%%%%%%%

For a general 2D metric, without imposing any further structure, the forms~(\ref{cf2DE}) and~(\ref{cf2DL}) appear to be the most convenient for our purposes. We denote quantities associated with the Lorentzian 2D part by the index $L$, while quantities associated with the Euclidean 2D parts are labeled by $\alpha,\beta,\ldots$ (there are $n/2-1$ such factors). We can then choose the null and spacelike frame vectors as
\begin{eqnarray}
&&\mathfrak{l} \ = \ \frac{1}{\Omega_L}(\partial_t \ + \ \partial_r)\, , \qquad \mathfrak{n}\ = \ \frac{1}{2\Omega_L}(-\partial_t \ + \ \partial_r)\, , \nonumber \\[2mm] 
&&\mathfrak{m}^{(\alpha1)}=\frac{1}{\sqrt{2}\Omega_\alpha}(\partial_{x_\alpha}+\ \partial_{y_\alpha})\, , \nonumber \\[2mm] 
&&\mathfrak{m}^{(\alpha2)} \ = \ \frac{1}{\sqrt{2}\Omega_\alpha}(\partial_{x_\alpha}- \ \partial_{y_\alpha})\, ,
\label{n-ada2+..+2Dv1}
\end{eqnarray}
The remaining 2D Euclidean factors are treated analogously, providing the required number of spacelike frame vectors, which we collectively denote by $i~\in~{\alpha_1,\alpha_2,\beta_1,\beta_2,\ldots}$.

Let us now determine the algebraic type of the $(2+\ldots+2)$ metrics. All such  geometries are necessarily algebraically special, since
\begin{eqnarray}
W_{0i0j} \ = \ W_{1i1j} \ \equiv \ 0\, .
\end{eqnarray}
Indeed, the Riemann tensor vanishes whenever its indices involve directions belonging to different factor subspaces. The remaining two contributions to the Weyl tensor in~(\ref{Weylten}) can, using (\ref{2DRiccit}), be reduced to terms proportional to $g_{\mu\nu}\mathfrak{l}^{\mu}\mathfrak{l}^{\nu}$ (and, analogously, to $g_{\mu\nu}\mathfrak{n}^{\mu}\mathfrak{n}^{\nu}$), which vanish by (\ref{n-ady}). The vanishing of $W_{0ijk}$ and $W_{1ijk}$ is even more direct, since each contribution to the corresponding Weyl components vanishes identically. Hence, all $(2+\ldots+2)$  geometries are at least of type II { (specifically, type $D$)}.

It remains to examine the boost-weight-zero components $W_{ijkl}$. For such a component to be nonzero, the spacelike frame indices must occur in two matching pairs coming from the same $\mathfrak{m}$: the first and third indices must correspond to the same spacelike direction, and likewise the second and fourth, or vice versa, but the two pairs must not coincide simultaneously. The two pairs may belong either to the same two-dimensional Euclidean factor or to two different factors. If both pairs belong to the same factor, we obtain 
\begin{eqnarray}
\label{pa}
W_{ijij} &=& R_{(s)}\left(\frac{1}{2}-\frac{1}{n-2}\right) \ + \ \frac{\sum R_{(a)}}{(n-1)(n-2)} \nonumber \\[2mm]
&=& -W_{ijji}\, ,
\end{eqnarray}
where $R_{(s)}$ denotes the Ricci scalar of the 2D subspace from which the $\mathfrak{m}$s are taken, while the sum runs over all $R$s, including $R_{(s)}$ and $R_L$. This projection can vanish if all spacelike Ricci scalars are equal and $R_L$ is correspondingly fixed by them. However, if the $\mathfrak{m}$s are taken from two different subspaces, we instead obtain
\begin{eqnarray}
\label{pb}
W_{ijij} &=& -\frac{R_{(s1)}+R_{(s2)}}{2(n-2)} \ + \ \frac{\sum R_{(a)}}{(n-1)(n-2)}\nonumber \\[2mm]
&=&-W_{ijji}\, ,
\end{eqnarray}
where $R_{(s1)}$ and $R_{(s2)}$ denote the Ricci scalars of the two corresponding 2D subspaces. The projections~(\ref{pa}) and~(\ref{pb}) cannot vanish simultaneously unless all $R$s vanish. The only exception occurs in four dimensions, where $R_{(1)}=-R_{(2)}$ is sufficient for both projections to vanish; this case is indeed conformally flat.

Before concluding this subsection, we note that, for the Schwarzschild-like coordinates associated with~(\ref{2Dcanonical}), one may alternatively choose the following frame:
\begin{eqnarray}
&&\mathfrak{n} \ = \ -\frac{1}{B_L}\partial_t\ + \ \partial_r\, , \quad \mathfrak{l} \ = \ \frac{1}{2}\partial_t \ + \ \frac{B_L}{2}\partial_r\, ,\nonumber \\[2mm]
&&\mathfrak{m}^{(\alpha1)} \ = \ \sqrt{\frac{B_\alpha}{2}}\partial_{x_\alpha} \ + \ \frac{1}{\sqrt{2B_\alpha}}\partial_{y_\alpha}\, , \nonumber \\[2mm]
&&\mathfrak{m}^{(\alpha2)} \ = \ \sqrt{\frac{B_\alpha}{2}}\partial_{x_\alpha}\ - \ \frac{1}{\sqrt{2B_\alpha}}\partial_{y_\alpha}\, ,
\label{n-ada2+..+2Dv2}
\end{eqnarray}
which is also a valid and convenient choice.

%%%%%%%%%%%%%%%%%%%%%%%%%%%%%%%%%
\subsection{$(2+m)$ metric structure}
%%%%%%%%%%%%%%%%%%%%%%%%%%%%%%%%%%

The simpler case corresponds to a {metric} in which the 2D sector is Lorentzian, while the $m$D part is Euclidean. The null vectors $\mathfrak{l}$ and
$\mathfrak{n}$ can be can be chosen {the same} as in previous subsection. We complete the  { vector} frame with the
spacelike vectors
\begin{equation}
\label{mwithoutr}
\mathfrak{m}^{(i)} \ = \ 1 \ + \ \frac{K\sum_{a=1}^m u_a^2}{4} \ \! \partial_{u_i}\, , 
\end{equation}
which are adapted to the metric~(\ref{Ecc}).

In the second case, the 2D part is Euclidean and the $m$D part is Lorentzian. We can again choose $\mathfrak{n}$ and $\mathfrak{l}$ as in~(\ref{n-ada2+..+2Dv2}), now with $B(r)=1-K_Lr^2$, and supplement them with the spacelike vectors
\begin{equation}
\label{mwithr}
\mathfrak{m}^{(i)} \ = \ \frac{1+\frac{1}{4}{\sum_{a=1}^m u_a^2}}{r} \ \! \partial_{u_i}\, ,
\end{equation}
for the remaining part of the $m$D subspace, adapted to~(\ref{Lcc}), together with two additional $\mathfrak{m}$ vectors for the 2D part, given by~(\ref{n-ada2+..+2Dv1}).

In both cases, for the same reason as for the $(2+\ldots+2)$ metrics, we have $W_{0i0j}=W_{1i1j}=0$. For the components $W_{0ijk}$ and $W_{1ijk}$, the two cases require slightly different arguments. If the 2D part is Lorentzian, these components vanish automatically. If the 2D part is Euclidean, all indices may belong to the Lorentzian $m$D part. Nevertheless, owing to the form of the Riemann tensor in~(\ref{mDccRiemann}), there is no term in~(\ref{Weylten}) that pairs an index associated with $\mathfrak{l}$ or $\mathfrak{n}$ with an index associated with $\mathfrak{m}$. These components therefore also vanish, and all $(2+m)$ metrics are at least of type II { (specifically, type $D$)}.

Let us now investigate whether further specialization is possible. For a nonzero $W_{ijkl}$, suppose first that the two index pairs are formed from the same $\mathfrak{m}$ vectors in the $m$D subspace. Then we obtain
\begin{eqnarray}
&&\mbox{\hspace{-12mm}}W_{ijij} \ = \ -W_{ijji} \nonumber \\[2mm]
&&\mbox{\hspace{-12mm}}= \ \frac{1}{m} \left( \frac{R_{(2D)}+R_{(mD)}}{m+1}-\frac{2R_{(mD)}}{m}+\frac{R_{(mD)}}{m-1}\right), 
\end{eqnarray}
which vanishes only if
\begin{eqnarray}
\label{2+mCF}
R_{(mD)} \ = \ -\frac{R_{(2D)} \ \! m(m-1)}{2}\, .
\end{eqnarray}
As will be shown in the next subsection, this condition corresponds precisely to the conformally flat case. If instead the two pairs of $\mathfrak{m}$ vectors are taken from different subspaces, the resulting expression again reduces to the same condition.

%%%%%%%%%%%%%%%%%%%%%%%%%%%%%%%%%%%%%%%
\subsection{$(m_1+m_2)$ metric structure}
%%%%%%%%%%%%%%%%%%%%%%%%%%%%%%%%%%%%%%%%%

First assume that $m_1,m_2\neq0,1$. We proceed as in the previous subsections. From the Lorentzian $m_1$D subspace, we choose $\mathfrak{l}$ and $\mathfrak{n}$ as in~(\ref{n-ada2+..+2Dv2}), with $B(r)=1-K_Lr^2$, together with the remaining $(m_1-2)$ spacelike vectors in the form~(\ref{mwithr}). From the Euclidean $m_2$D subspace, we choose the remaining spacelike vectors in the form (\ref{mwithoutr}), with $K\equiv K_E$.

For the same reason as in the previous subsection, we have $W_{0i0j}=W_{1i1j}=0$ and $W_{0ijk}=W_{1ijk}=0$. Hence, $(m_1+m_2)$ type of metrics with constant-curvature subspaces are at least of type II { (specifically, type $D$)}. We next examine whether further specialization is possible. For $W_{ijkl}$ to be nonzero, the two index pairs must be formed from two $\mathfrak{m}$ vectors belonging to different subspaces, which is always possible. In this case, we obtain
\begin{eqnarray}
W_{ijij} \ &=& \ -W_{ijji} \nonumber \\[2mm]
&=&\frac{1}{n-2} \left(\frac{R_{(m_1D)}+R_{(m_2D)}}{n-1}\right)\nonumber \\[2mm]
&-&\frac{1}{n-2} \left(\frac{R_{(m_1D)}}{m_1} + \frac{R_{(m_2D)}}{m_2} \right)\, . 
\end{eqnarray}
This component vanishes if
\begin{eqnarray}
\label{m1+m2CF}
R_{(m_2D)} \ = \ -\frac{R_{(m_1D)} \ \! m_2(m_2-1)}{m_1(m_1-1)}\, ,
\end{eqnarray}
which corresponds precisely to the conformally flat case. Indeed, substituting this into the general expression for the Weyl tensor components (which can be nonzero only if either all indices are from one subspace or two from one, another two from second), we get exactly the same condition for vanishing.

Specifically in the $(m+1)$  case, there is no need to construct the explicit {vector} frame, since it is clear that the Weyl tensor vanishes identically. For most components, all terms in~(\ref{Weylten}) vanish automatically, while for the remaining components, involving an even number of 1D indices, the nonzero terms cancel pairwise. Hence, these geometries are conformally flat. The $n$D constant-curvature metrics, corresponding to anti-de Sitter, de Sitter, or Minkowski ST, are likewise well known to be conformally flat. 

%\textbf{[Obecně $m+1$ a $n$D případ většinou řešíme zvlášť. Možná lepší (alespoň někde) sjednotit.]}

%%%%%%%%%%%%%%%%%%%%%%%%%%%%%%
\section{Conformal relations for $(m +1)$ metrics}
\label{ACEST}
%%%%%%%%%%%%%%%%%%%%%%%%%%%%%%

In this section, we consider $(m+1)$ DP metrics in which the $m$D factor has constant curvature. As established in the previous section, these ST are conformally flat. Here, we explicitly construct the conformal factors that map them to flat ST, as well as those relating them to other LG solutions.

First, we recall the transformation laws for the Ricci tensor and Ricci scalar under the Weyl transformation~(\ref{WTr}), see e.g., Ref.~\cite{Straumann:2013}
\begin{eqnarray}
\tilde R_{\mu \nu} \ &=& \ R_{\mu \nu} \ - \ (n-2)(\nabla_\mu  \nabla_\nu \omega-\nabla_\mu \omega \nabla_\nu \omega) \nonumber \\[2mm]
&&- \  \big(\Box \omega \ + \ (n-2) \partial^\lambda \omega \partial_\lambda \omega\big)\ \! g_{\mu \nu}\, ,
\label{RicciTWeylT}
\end{eqnarray}
and 
\begin{eqnarray}
\label{RicciSWeylT}
\tilde{R}\ &=& \ e^{-2\omega}\big[R \ - \ 2(n-1)\Box \omega\big] \nonumber \\[2mm]
&-& \ e^{-2\omega}(n-2)(n-1)\partial^\lambda \omega \partial_\lambda \omega\, .
\end{eqnarray}

%%%%%%%%%%%%%%%%%%%%%%%%%%%%%%%%%%%%%%
\subsection{Euclidean $m$D part}
%%%%%%%%%%%%%%%%%%%%%%%%%%%%%%%%%%%%%%%

Now assume metric of the form 
\begin{eqnarray}
ds^2 \ = \-dt^2 \ + \ d\Sigma^2_{m, E, K}\, ,
\end{eqnarray}
where $d\Sigma^2_{m,E,K}$ is given by~(\ref{Ecc}), so that more explicitly
\begin{eqnarray}
&&g_{tt} \ = \ -1\, , \qquad g_{aa} \ = \ \frac{16}{(4+K\sum_{i=1}^m u_i^2)^2}\, , \nonumber \\[2mm] 
&&R_{tt} \ = \ 0\, , \qquad R_{ab} \ = \ K(m-1)g_{ab}\, ,\nonumber \\[2mm]
&&R \ = \ Km(m-1)\, .
\end{eqnarray}
Under the Weyl transformation with $\omega=\omega(t)$ {ansatz}, these quantities transform as
\begin{eqnarray}
&&\mbox{\hspace{-10mm}}
\tilde{g}_{\mu\nu}
\ = \ 
e^{2\omega}g_{\mu\nu}\, , \nonumber \\[2mm]
&&\mbox{\hspace{-10mm}}\tilde{R}_{tt} \ = \ -m\omega''\, , \nonumber \\[2mm]
&&\mbox{\hspace{-10mm}}\tilde R_{ab} \ = \ \frac{16\big[K(m-1)+\omega''+(m-1)(\omega')^2\big]}{(4+K\sum_{i=1}^m u_i^2)^2}\, , \nonumber \\[2mm]
&&\mbox{\hspace{-10mm}}\tilde{R} \ = \ m[K(m-1)+2\omega''+(m-1)(\omega')^2]e^{-2\omega}\, .
\end{eqnarray}
The LG equations, $\tilde G_{\mu \nu}+\Lambda\tilde g_{\mu \nu}=0$, then reduce to
\begin{equation}
\label{t-komp}
\frac{m(m-1)[K+(\omega')^2]}{2} \ - \ \Lambda e^{2\omega} \  = \ 0\, ,
\end{equation}
for the $tt$-component, and
\begin{eqnarray}
&&\mbox{\hspace{-13mm}}\omega''(1-m) \ + \ (\omega')^2 \frac{(m-1)(2-m)}{2} \nonumber \\[2mm]
&&\mbox{\hspace{3mm}}+ \ \frac{K(m-1)(2-m)}{2} \ + \ \Lambda e^{2\omega} \ = \ 0\, ,
\label{ab-komp}
\end{eqnarray}
for the $ab$-components, where the equation has been traced over the $m$D part and divided by $m$.

We now choose the normalization of the metric such that
$K\in \{-1,0,1\}$. For each value of $K$ and nonzero $\Lambda$, equation~(\ref{t-komp}) admits solutions, either with $\omega$ constant or with $\omega$ a non-constant function of $t$. The constant solution is incompatible with the second equation~(\ref{ab-komp}), whereas the non-constant solution of~(\ref{t-komp}) satisfies~(\ref{ab-komp}) automatically. The resulting solutions are [using the notation of~(\ref{WTr})]
\begin{eqnarray}
&&\mbox{\hspace{-5mm}}\Omega^2 \ = \ \frac{m(m-1)}{2\Lambda f_K^2(t)}\, , 
 \label{f_iE}
\end{eqnarray}
with
\begin{eqnarray}
&&f_{-1} \ = \ \sinh(C-t)\, ,\quad f_1 \ = \ \cos(C-t)\, , \nonumber \\[2mm]
&&f_0 \ = \ C \ - \ t\, ,
\label{B9.cc}
\end{eqnarray}
where $C$ is arbitrary constant. We note that $f_1$ can be also chosen as $\sin(C-t)$. Since $\Omega^2$ must be positive, so must be $\Lambda$. For $\Lambda=0$ there is obviously constant solution for $K=0$ case (since this is Minkowski already) and for $K=-1$ case there exist a linear one: $\omega(t)=\pm t+C$.

%%%%%%%%%%%%%%%%%%%%%%%%%%%%%%%%%%%%%%
\subsection{Lorentzian $m$D part}
%%%%%%%%%%%%%%%%%%%%%%%%%%%%%%%%%%%%%

Here we consider the metric
\begin{eqnarray}
ds^2 \ = \ dx^2 \ + \ d\Sigma^2_{m,L,K}\, ,
\end{eqnarray}
with the Lorentzian constant-curvature metric specified by~(\ref{Lcc}),  i.e.
\begin{eqnarray} 
&&g_{tt} \ = \ -(1-Kr)\, , \quad g_{rr} \ = \ (1-Kr)^{-1}\, , \nonumber \\[2mm]
&&g_{aa} \ = \ \frac{16r^2}{\left(4+\sum_{i=1}^m u_i^2\right)^2}\, , \quad g_{xx} \ = \ 1\, ,  \nonumber 
\\[2mm] 
&&R_{ab} \ = \ K(m-1)g_{ab}\, ,  \quad R_{xx} \ = \ 0\, , \nonumber \\[2mm]
&&R \ = \ m(m-1)K\, .
\end{eqnarray}
We now perform the Weyl transformation  with $\omega=\omega(x)$ and transform the Ricci tensor and scalar according to~(\ref{RicciTWeylT}) and~(\ref{RicciSWeylT}). This gives
\begin{eqnarray}
&&\mbox{\hspace{-13mm}}\tilde{R}_{ab} \ = \ \big[K(m-1)-\omega''-(m-1)(\omega')^2\big]g_{ab}\, , \nonumber \\[2mm]
&&\mbox{\hspace{-13mm}} \tilde{R}_{xx} \ = \ -m\omega''\, , \nonumber \\[2mm] 
&&\mbox{\hspace{-13mm}}\tilde{R} \ = \ m\big[K(m-1)-2\omega''-(m-1)(\omega')^2\big]e^{-2\omega}\, .
\end{eqnarray}
The corresponding transformed LG equations are 
\begin{equation}
\frac{m(m-1)[(\omega')^2-K]}{2} \ + \ \Lambda e^{2\omega} \ = \ 0\, ,
\end{equation}
obtained from the $xx$-component and 
\begin{eqnarray}
&&\mbox{\hspace{-10mm}}\omega''(m-1) \ + \ (\omega')^2 \ \! \frac{(m-1)(m-2)}{2}\nonumber \\[2mm]
&&\mbox{\hspace{3mm}}+ \ \frac{K(m-1)(2-m)}{2} \ + \ \Lambda e^{2\omega} \ = \ 0\, ,
\end{eqnarray}
from the $ab$-components, where the equation has been
traced over the $m$D part and divided by $m$. 

As in the Euclidean case, it is sufficient to first solve the $xx$-component. For normalized $K$ and nonzero $\Lambda$, this equation admits one constant and one non-constant solution. The constant solution, however, fails to satisfy the $ab$-component, whereas the non-constant solution satisfies both components. The resulting solutions are
\begin{eqnarray}
\Omega^2 \ = \ \frac{m(1-m)}{2\Lambda f_K^2(x)}\, , \quad 
\end{eqnarray}
with
\begin{eqnarray}
&&f_1 \ = \ \sinh(C-x)\, , \quad f_{-1} \ = \ \cos(C-x)\, , \nonumber \\[2mm]
&&f_0 \ = \ C \ - \ x\, ,
\label{f_iL}
\end{eqnarray}
with $C$ being an arbitrary constant. Note that $f_1$ and $f_{-1}$ are interchanged relative to~(\ref{B9.cc}). Equivalently, one may choose $f_{-1}=\sin(C-x)$. Since $\Omega^2$ must be positive, $\Lambda$ must be negative. In the absence of $\Lambda$, there is a constant solution for $K=0$, since the original metric is already Minkowski, while for $K=1$ there is a linear solution: $\omega(x)=\pm x+C$. \\

%\textbf{[Vypadá to, že (\ref{f_iE}) transformují do de Sittera a (\ref{f_iL}) do anti de Sittera, ještě ověřím.]}

%%%%%%%%%%%%%%%%%%%%%%%%%%%%%%
\section{Explicit non-trivial solutions of $\Box R=0$ in 2D}
\label{Abox}
%%%%%%%%%%%%%%%%%%%%%%%%%%%%%%

In this appendix and the following one, we consider the equation
\begin{eqnarray}
\Box_{(2D)} R_{(2D)} \ + \ C R_{(2D)} \ = \ 0\, ,
\end{eqnarray}
which arose in Section~\ref{QG} in the context of QG, and investigate its possible solutions. In the present appendix, we focus on the case $C=0$. We treat the Lorentzian and Euclidean cases separately. The most convenient metric forms for this investigation appear to be Eqs.~(\ref{cf2DE}) and~(\ref{cf2DL}). In these forms, the equation under consideration reduces to
\begin{equation}
\Box_{(2D)} R_L \ \equiv \ ( \partial_r^2 R_L \ - \ \partial_t^2 R_L) e^{-2\omega(t,r)} \ = \ 0\, ,
\end{equation}
and 
\begin{equation}
\Box_{(2D)} R_E \ \equiv \ ( \partial_x^2 R_E \ + \ \partial_y^2 R_E) e^{-2\omega(x,y)} \ = \ 0\, .
\end{equation}
Using the explicit expressions~(\ref{ER}) and~(\ref{LR}) for the corresponding Ricci scalars, equation $\Box R_{L,E}=0$ takes the form

\begin{widetext}
\begin{eqnarray}
&&\mbox{\hspace{-15mm}}\Biggl \{ - \frac{\partial^4 \omega}{\partial r^4} \ - \ \frac{\partial^4 \omega}{\partial t^4} \ + \ 2\frac{\partial^4 \omega}{\partial r^2 t^2} \ + \ 4 \left[ \frac{\partial \omega}{\partial r} \frac{\partial^3 \omega}{\partial r^3} \ - \ \frac{\partial \omega}{\partial t} \frac{\partial^3 \omega}{\partial r^2 \partial t} \ - \ \frac{\partial \omega}{\partial r} \frac{\partial^3 \omega}{\partial r \partial t^2} \ + \ \frac{\partial \omega}{\partial t} \frac{\partial^3 \omega}{\partial t^3}\right]\nonumber\\[2mm]
&&\mbox{\hspace{5mm}}+ \ 2\left[\frac{\partial^2 \omega}{\partial r^2} \ - \ \frac{\partial^2 \omega}{\partial t^2} \right] \left[ 2\left( \frac{\partial \omega}{\partial t}\right)^2 \ - \ 2\left( \frac{\partial \omega}{\partial r}\right)^2 \ +  \ \frac{\partial^2 \omega}{\partial r^2} \ - \  \frac{\partial^2 \omega}{\partial t^2} \right] \Biggr \}2e^{-4\omega} \ = \ 0\, ,  
 \label{LPDE}
\end{eqnarray}
for the Lorentzian version and

\begin{eqnarray}
&&\mbox{\hspace{-15mm}}\Biggl \{ - \frac{\partial^4 \omega}{\partial x^4} \ - \ \frac{\partial^4 \omega}{\partial y^4} \ - \ 2\frac{\partial^4 \omega}{\partial x^2 y^2} \ + \ 4 \left[ \frac{\partial \omega}{\partial x} \frac{\partial^3 \omega}{\partial x^3} \ + \ \frac{\partial \omega}{\partial y} \frac{\partial^3 \omega}{\partial x^2 \partial y} \ + \ \frac{\partial \omega}{\partial x} \frac{\partial^3 \omega}{\partial x \partial y^2} \ + \ \frac{\partial \omega}{\partial y} \frac{\partial^3 \omega}{\partial y^3}\right]\nonumber\\[2mm]
&&\mbox{\hspace{5mm}}
+2\left[\frac{\partial^2 \omega}{\partial x^2} \ + \  \frac{\partial^2 \omega}{\partial y^2} \right] \left[ -2\left( \frac{\partial \omega}{\partial x}\right)^2 \ - \ 2\left( \frac{\partial \omega}{\partial y}\right)^2 \ + \ \frac{\partial^2 \omega}{\partial x^2} \ + \  \frac{\partial^2 \omega}{\partial y^2} \right] \Biggr \}2e^{-4\omega}\ = \ 0\, ,
 \label{EPDE}
\end{eqnarray} 
for the Euclidean version.
\end{widetext}

While we have not been able to obtain a general solution to these equations, we can nevertheless exhibit several nontrivial examples. In particular, consider
\begin{equation}
\omega(r,t) \ = \ -\frac{3}{2}\ln(c_1r \ + \ c_2t \ + \ c_3)\, ,
\end{equation}
in the Lorentzian case, and, analogously
\begin{equation}
\omega(x,y) \ = \ -\frac{3}{2}\ln(c_1x \ + \ c_2y \ + \ c_3)\, ,
\end{equation}
in the Euclidean case. If the coefficient $3/2$ is replaced by $1$, these expressions instead yield solutions with constant Ricci scalar. The same is true when $c_1=c_2=0$, as well as for various other choices of $\omega$ that we do not list here, since they merely provide more complicated representations of 2D constant-curvature spaces.

An alternative approach is to work in Schwarzschild-like coordinates. For the Lorentzian case, we take
\begin{equation}
ds^2_L \ = \ -B(r,t)dt^2 \ + \ B(r,t)^{-1}dr^2\, ,
\end{equation}
while in the Euclidean case we consider
\begin{equation}
ds^2_E \ = \ B(x,y)dx^2 \ + \ B(x, y)^{-1}dy^2\, ,
\end{equation}
The corresponding Ricci scalars are
\begin{equation}
R_L \ = \ \frac{-B^3\partial_r^2B \ + \ 2(\partial_t B)^2 \ - \ B\partial^2_tB}{B^3}\, ,
\end{equation}
and
\begin{equation}
R_E \ = \ \frac{-B^3\partial_y^2B \ - \ 2(\partial_x B)^2 \ + \ B\partial^2_xB}{B^3}\, .
\end{equation}
For these parametrizations, we find the following nontrivial examples:
\begin{eqnarray}
B(t,r) \ = \ (c_1r+c_2)^{\frac{3}{2}}\, ,
\end{eqnarray}
or
\begin{eqnarray}
B(t,r) \ = \ (c_1t+c_2)^{-\frac{3}{2}}\, ,
\end{eqnarray}
for Lorentzian case and
\begin{equation}
B(x,y) \ = \ e^{c_1x+c_2}\, , 
\end{equation}
or
\begin{equation}
B(x,y) \ = \ e^{c_1y+c_2}\, .
\end{equation}
for Euclidean case. 

%{\color{black}\textbf{[Dost možná, jde o speciální případy těch výše, zkusit převod souřadnic.]}}

%%%%%%%%%%%%%%%%%%%%%%%%%%%%%%%%%%%%%%%%%%%%%%%%%
\section{Explicit non-trivial solutions of $\Box R+CR=0$ in 2D}
\label{Afullbox}
%%%%%%%%%%%%%%%%%%%%%%%%%%%%%%%%%%%%%%%%%%%%%%%%%%%

In this appendix, we establish the existence of non-trivial solutions to $\Box R + CR = 0$ with $C\neq0$ in 2D and obtain these
solutions in closed form, up to quadrature.\\

%%%%%%%%%%%%%%%%%%%%%%%%%%%%%%%%%%%%%%%%%%%%%%%%%
\subsection{Spatial coordinate ansatz \label{D.1.ch}}
%%%%%%%%%%%%%%%%%%%%%%%%%%%%%%%%%%%%%%%%%%%%%%%%%

Consider first $\omega(r,t)=f(r)$ in the Lorentzian case, or
$\omega(x,y)=f(x)$ in the Euclidean case. In both cases [cf. also Eqs.~(\ref{ER}) and~(\ref{LR})]
\begin{eqnarray}
R \ = \ -2e^{-2f}f''\, ,
\qquad
\Box R \ = \ e^{-2f}R''\, .
\label{5.df}
\end{eqnarray}
Hence $\Box R+CR=0$ is equivalent to
\begin{eqnarray}
R''\ + \ Ce^{2f}R \ = \ 0\, . \label{1.af}
\end{eqnarray}
Since $e^{2f}R=-2f''$, Eq.~(\ref{1.af}) becomes
\begin{eqnarray}
R''\ - \ 2Cf''\ = \ 0\, .
\end{eqnarray}
Therefore
\begin{eqnarray}
R \ - \ 2Cf \ = \ Ar \ + \ B\,.
\end{eqnarray}
Here, $A$ and $B$ are arbitrary constants. In the following, for simplicity, we restrict our attention to the sector of solutions with $A=B=0$. The equation then reduces to the simple form
\begin{eqnarray}
R \ = \ 2Cf\, ,
\end{eqnarray}
which might be equivalently rewritten as
\begin{eqnarray}
-2e^{-2f}f''\ = \ 2Cf\, .
\end{eqnarray}
Thus $f$ satisfies the autonomous second-order differential equation
\begin{eqnarray}
f''\ + \ Ce^{2f}f \ =\ 0. \label{2.gh}
\end{eqnarray}
A standard approach to solving these equations is to multiply
Eq.~(\ref{2.gh}) by $f'$ and integrate once, yielding
\begin{eqnarray}
\frac12(f')^2
\ + \ \frac{C}{4}e^{2f}(2f-1)
\ = \ E\, , \label{3.sd}
\end{eqnarray}
where $E$ is an arbitrary integration constant. Consequently,
\begin{eqnarray}
f'
\ = \ 
\pm
\sqrt{
2E\ - \ \frac{C}{2}e^{2f}(2f-1)
}\, .
\end{eqnarray}
The solution is therefore given exactly by the quadrature
\begin{eqnarray}
s \ - \ s_0
\ = \ 
\pm
\int_{f(s_0)}^{f(s)}
\frac{du}{
\sqrt{
2E \ - \ \frac{C}{2}e^{2u}(2u-1)
}
}\, . \label{4.kl}
\end{eqnarray}
%

%%%%%%%%%%%%%%%%%%%%%%%%%%%%%%%%%%
\subsubsection{The case \(C>0\)}
%%%%%%%%%%%%%%%%%%%%%%%%%%%%%%%%%%%%

Note that the function
\begin{eqnarray}
V_+(u) \ = \ \frac{C}{4}e^{2u}(2u-1)\, ,
\end{eqnarray}
has a strict minimum at $u=0$, with
\begin{eqnarray}
V_+(0) \ = \ -\frac{C}{4}\, ,
\end{eqnarray}
and 
the function satisfies
\begin{eqnarray}
\lim_{u\to-\infty}V_+(u) \ = \ 0^{-}\, ,\quad
\lim_{u\to+\infty}V_+(u) \ = \ +\infty\, .~~~
\end{eqnarray}
See Tab.~\ref{tab:solution_classification} for classification of solution types.
%
%\begin{widetext}
%\begin{center}
\begin{table*}[ht]
\centering
\renewcommand{\arraystretch}{1.3}
\begin{tabular}{c|c}
\hline\hline
\textbf{Energy} & \textbf{Solution type} \\
\hline\hline
\\[-3mm]
$E<-\dfrac{C}{4}$
&
No real solution
\\[1.5mm]
$E=-\dfrac{C}{4}$
&
$f\equiv0$
\\[1.5mm]
$-\dfrac{C}{4}<E<0$~
&
~Periodic solution
\\[1.5mm]
$E=0$
&
Global bounce
\\[1.5mm]
$E>0$
&
Global bounce
\\
\hline
\end{tabular}
\caption{Classification of solutions according to the value of the
 parameter $E$.}
\label{tab:solution_classification}
\end{table*}
%\end{center}
%\end{widetext}
%
In particular, for the periodic branch $-\frac{C}{4}<E<0$,
there are two finite turning points $u_-<0<u_+$ [determined by the equation $V_+(u) = E$, see Eq.~(\ref{3.sd})], and the solution oscillates periodically between $u_-$ and $u_+$ with the period 
\begin{eqnarray}
T(E)
\ = \ 
2\int_{u_-}^{u_+}
\frac{du}{
\sqrt{
2E \ - \ \frac{C}{2}e^{2u}(2u-1)
}
}\, . \label{5.kk}
\end{eqnarray}
For $E\geq0$, there is only one finite turning point $u_+$. The solution is defined
implicitly by
\begin{eqnarray}
|s-s_0| \ = \ 
\int_{f(s)}^{u_+}
\frac{du}{
\sqrt{
2E \  - \ \dfrac{C}{2}e^{2u}(2u-1)
}
}\, , \label{3.kl}
\end{eqnarray}
and it satisfies
\begin{eqnarray}
&&f(s_0) \ = \ u_+\, ,\qquad f'(s_0) \ = \ 0\, ,
\nonumber \\[2mm]
&&f(s) \ \to \ -\infty
\quad\text{as}\quad s \ \to \ \pm\infty\, .
\end{eqnarray}
Thus it is a global non-periodic bounce solution.

In all aforementioned cases
$R(s)  =  2Cf(s)$,
a direct substitution into the original equation  gives
\begin{eqnarray}
\Box R \ + \ CR
\ = \ 
e^{-2f}(2Cf'') \ + \ 2C^2f \ = \ 0\, ,
\end{eqnarray}
which, in turn, confirms that the solutions $f$ represent a specific class of solutions of our 2D equation.

%%%%%%%%%%%%%%%%%%%%%%%%%%%%%%%%
\subsubsection{The case \(C<0\)}
%%%%%%%%%%%%%%%%%%%%%%%%%%%%%%%

In this case, we  set $C=-\kappa^2$ with $\kappa>0$.
The corresponding reduced equation~(\ref{2.gh}) becomes
\begin{eqnarray}
f''\ - \ \kappa^2e^{2f}f \ = \ 0, \label{4.fg}
\end{eqnarray}
with first integral
\begin{eqnarray}
&&\frac12(f')^2 \ + \ V_-(f) \ = \ E\,, \nonumber \\[2mm]
&&V_-(f) \ = \ \frac{\kappa^2}{4}e^{2f}(1-2f)\, . ~~\label{5.sd}
\end{eqnarray}
Here
\begin{eqnarray}
&&V_-(0) \ = \ \frac{\kappa^2}{4}\, ,\qquad
\lim_{f\to-\infty}V_-(f) \ = \ 0^{+}\, ,\nonumber \\[2mm]
&&\lim_{f\to+\infty}V_-(f) \ = \ -\infty\, .
\end{eqnarray}
The possible branches are represented in Tab.~\ref{tab:energy_classification-1}. 

\begin{table*}[t]
\centering
\renewcommand{\arraystretch}{1.35}
\begin{tabular}{c|c}
\hline\hline
\textbf{Energy} & \textbf{Solution type} \\
\hline\hline
\\[-3mm]
$E\ < \ 0$ & Unbounded solution; $f\to+\infty$ at finite $s$ \\[2mm]
$E \ = \ 0$ & Unbounded solution with finite-distance blow-up \\[2mm]
$0\ < \ E \ < \ \dfrac{\kappa^2}{4}$~~&~~Left global bounce or right finite-distance blow-up \\[2mm]
$E \ = \ \dfrac{\kappa^2}{4}$ & Constant solution $f=0$ and separatrix branches \\[2mm]
$E \ > \ \dfrac{\kappa^2}{4}$ & Unbounded branch with finite-distance blow-up \\[2mm]
\hline
\end{tabular}
\caption{Classification of solution behaviour according to the value of the
energy parameter $E$.}
\label{tab:energy_classification-1}
\end{table*}

For $0<E<\frac{\kappa^2}{4}$, the equation
$V_-(f)=E$
has two roots $u_-<0<u_+$ (explicit solutions can be explained in terms of Lambert $W$ functions). 

The branch $f\leq u_-$
has one finite turning point and satisfies
\begin{eqnarray}
f(s)\ \to \ -\infty
\quad\text{as}\quad s\ \to \ \pm\infty\,.
\end{eqnarray}
Hence it is a global bounce and it is given implicitly by
\begin{eqnarray}
|s-s_0| \ = \ 
\int_{f(s)}^{u_-}
\frac{du}{
\sqrt{
2E\ - \ \dfrac{\kappa^2}{2}e^{2u}(1-2u)
}
}\, . \label{6.uj}
\end{eqnarray}

The other allowed branch
$f\geq u_+$, reaches $f=+\infty$ at finite $s$.

Consequently, for $C<0$, the spatial one-variable reduction has no periodic
solutions, but it does contain global non-periodic bounce solutions for
\begin{eqnarray}
0 \ < \ E \ < \ \frac{|C|}{4}\, .
\end{eqnarray}

%%%%%%%%%%%%%%%%%%%%%%%%%%%%%%%%%%%%%%%%%%%%%%%%%
\subsection{Time ansatz in the Lorentzian case}
%%%%%%%%%%%%%%%%%%%%%%%%%%%%%%%%%%%%%%%%%%%%%%%%%

In the Euclidean case, there are two interchangeable spatial coordinates; hence, the two choices of coordinate ansatz lead to the same type of solution. This is no longer the case for a Lorentzian metric. While for the spatial ansatz we have still~(\ref{5.df}), for the time coordinate ansatz $\omega(r,t)=f(t)$, we obtain
\begin{eqnarray}
R \  = \ 2e^{-2f}f''\, ,
\qquad
\Box R \ = \ -e^{-2f}R''\, .
\end{eqnarray}
Therefore $\Box R \ + \ CR \ = \ 0$ becomes
\begin{eqnarray}
R''\ - \ Ce^{2f}R \ = \ 0\, .
\end{eqnarray}
As in~\ref{D.1.ch}, we can select
$R  =  2Cf$, which leads to
\begin{eqnarray}
f''\ - \ Ce^{2f}f \ = \ 0\, . \label{10.hh}
\end{eqnarray}
Consequently, when
\begin{eqnarray}
C<0,\qquad C \ = \ -\kappa^2\, ,
\end{eqnarray}
eq.~(\ref{10.hh}) becomes
\begin{eqnarray}
f''\ + \ \kappa^2e^{2f}f \ = \ 0\, ,
\end{eqnarray}
which is exactly the oscillatory equation discussed above with
$C$ replaced by $\kappa^2=|C|$. Hence, for every
\begin{eqnarray}
-\frac{|C|}{4} \ < \ E \ < \ 0\, ,
\end{eqnarray}
there are global periodic solutions
\begin{eqnarray}
\omega(r,t) \ = \ f(t)\, ,
\end{eqnarray}
where $f$ is given by
\begin{eqnarray}
t \ - \ t_0
\ = \ 
\pm
\int_{f(t_0)}^{f(t)}
\frac{du}{
\sqrt{
2E \ - \ \frac{|C|}{2}e^{2u}(2u-1)
}
}\, . ~~~\label{11/as}
\end{eqnarray}
Thus the periodic one-coordinate solutions can be summarized as\\[-1mm]
\begin{eqnarray*}
\begin{array}{c|c|c}
\text{Signature} & \text{Dependence} & \text{Periodic sector}\\
\hline\\[-3mm]
(-+) & \omega  =  f(r) & C \ > \ 0\\[1mm]
(-+) & \omega  =  f(t) & C \ < \ 0\\[1mm]
(++) & \omega  =  f(x)\ \text{or}\ f(y) & C \ > \ 0
\end{array}
\end{eqnarray*}\\[-1mm]
For the Euclidean metric and \(C<0\), a one-coordinate ansatz still gives
the exact equation (6), but the effective potential $V_{-}$ is nonconfining with a bounce solution.
Therefore this particular reduction yields no nonconstant periodic global
solution.

%%%%%%%%%%%%%%%%%%%%%%%%%%%%%%
\section{4D and $(2+\ldots+2)$ solutions with symmetry sharing EM field}
\label{AEM}
%%%%%%%%%%%%%%%%%%%%%%%%%%%%%%

Here we consider the inclusion of an EM field into our models. In particular, we introduce an EM field compatible with the DP structure in all 4D cases considered, as well as in the general $(2+\ldots+2)$ configurations of QG. This analysis extends the results of Ref.~\cite{previous}, where the $(2+2)$ case was studied in WCG.

The general EM stress-energy tensor is
\begin{equation}
\label{TEM}
T_{\mu \nu}^{EM} \ = \ \frac{1}{e^2}\left(F_{\mu \lambda}F_\nu^\lambda \ - \ \frac{1}{4}g_{\mu \nu} F_{\alpha \beta} F^{\alpha \beta}\right),
\end{equation}
where $e$ denotes the EM coupling constant.

We begin with the $(2+2)$ case, i.e., allowing $F_{01}$ and $F_{23}$ to be the only non-vanishing components of the EM field strength. In general, these components may depend on the coordinates of their respective two-dimensional parts. Using the result of~\cite{previous}, we have
\begin{equation}
\label{2+2EM}
T_{ij} \ =  \ \frac{1}{2e^2} \left[(F^{01})^2g_{(1)} \ - \ (F^{23})^2g_{(2)}\right]g_{ij}\, .
\end{equation}
Here $g_{(1)}$ and $g_{(1)}$ represent the determinants of the metric of the respective 2D subspaces.
Thus, the EM stress-energy tensor is proportional to the metric on each factor and therefore contributes to the right-hand side of the FEs without altering their tensorial structure. In particular, for $2\alpha+\beta\neq 0$, KVFs remain present.

The Maxwell equations (MEs) take the form 
\begin{equation}
\label{ME}
\partial_\mu(\sqrt{\vert g \vert}F^{\mu \nu}) \ = \ 0\, ,
\end{equation}
which for assumptions and 2D metrics form (\ref{2Dcanonical}) means both EM field components are constant. Consequently, the components of $T_{\mu\nu}$ in~(\ref{2+2EM}) are of the form of a constant multiplying the corresponding metric tensor components. The EM field therefore contributes an additional constant term to the right-hand side of~(\ref{4D2+2full}) or~(\ref{4D2+2bezb3}).

%\textbf{[Možná napsat explicitně. "Povolené" teorie, vztah k \cite{previous}.]}

For the $(3+1)$ case, three independent components of the EM field strength may, in principle, be non-zero within the 3D part.  However, the constant-curvature metrics~(\ref{Ecc}) and~(\ref{Lcc}), i.e., the coordinate representations considered here, are diagonal. Consequently, the off-diagonal components of the first term in the stress-energy tensor~(\ref{TEM}) must vanish. This implies that at most one of the three independent components can be non-zero. To say more, we must look at the MEs  for the  particular metric. 

Consider first the case in which the 1D part represents the time direction, $x^0=t$, while the 3D part is Euclidean and described by the metric~(\ref{Ecc}). Suppose, without loss of generality, that the only potentially non-vanishing component is $F_{12}$. The MEs~(\ref{ME}) then imply
\begin{eqnarray}
\partial_1\left(\sqrt{\vert g\vert}F^{12}\right)
\ = \
\partial_2\left(\sqrt{\vert g\vert}F^{12}\right)
\ = \ 0, .
\end{eqnarray}
At the same time, the Bianchi identities (BIs)
\begin{equation}
\label{BI}
\partial_\lambda F_{\mu \nu} \ + \ \partial_\mu F_{\nu \lambda} \ + \ \partial_\nu F_{\lambda \mu} \ = \ 0\, ,
\end{equation}
must also be satisfied.  Choosing $\mu=1$, $\nu=2$, and $\lambda=3$ gives
\begin{equation}
\partial_3 F_{12} \ = \ 0\, .
\end{equation}
For the metric~(\ref{Ecc}), this condition is incompatible with the MEs unless $F_{12}=0$ or $K=0$. In the latter case, $F_{12}$ may be a non-zero constant. Since $K=0$ corresponds to Minkowski ST, this case is already contained in the 4D constant-curvature analysis discussed below.

Now let the 1D part correspond to the spatial coordinate $x^3$, and let the 3D part be Lorentzian. The latter is then described by the metric~(\ref{Lcc}). For a single non-vanishing component, $F_{12}$, $F_{01}$, or $F_{02}$, the MEs alone admit solutions. However, the left-hand side of the FEs has the form of the metric components multiplied by constants, and hence the same structure must hold for $T_{\mu\nu}$. This is not the case for these EM configurations, except for the special case in which $F_{12}$ is non-zero and $K=0$. For $K=0$, however,~(\ref{nDcc}) and~(\ref{n-1+1Dcc}) imply that the left-hand side of the FEs vanishes identically. Consequently, $T_{\mu\nu}$ must also vanish, which again requires $F_{12}=0$. Thus, there is no non-trivial EM field compatible with these $(3+1)$ direct-product STs.

It remains to consider the 4D constant-curvature case. Here,~(\ref{nDcc}) implies that $T_{\mu\nu}$ must vanish, since the gravitational part of the FEs vanishes identically. The requirement that the off-diagonal components of $T_{\mu\nu}$ vanish permits, in principle, the following pairs of non-vanishing field-strength components:
\begin{eqnarray}
(F^{01},F^{23})\, , \qquad (F^{02},F^{13})\, , \qquad (F^{03},F^{12})\, .
\end{eqnarray}
For each pair, the vanishing of the corresponding off-diagonal components requires the squares of the two non-zero field-strength components to be equal. However, using the metric~(\ref{Lcc}), this condition cannot be satisfied for non-zero field strengths. Hence, no non-trivial EM field is compatible with the 4D constant-curvature STs.

Two further remarks concerning the EM field are in order. First, Yang--Mills fields can be incorporated in an analogous manner; see~\cite{previous}. Second, the procedure leading to the form of the EM stress-energy tensor in the $(2+2)$ case extends straightforwardly to an arbitrary number of 2D subspaces. In general, we obtain
\begin{equation}
T_{ij(k)}=\frac{1}{2e^2}\left[(F^{(k)})^2g_{(k)}-\sum_{l \neq k}(F^{(l)})^2g_{(l)}\right]g_{ij(k)}\, ,
\end{equation}
where $F^{(k)}$ denotes the EM field-strength component associated with the $k$th 2D sector, in the same sense as in~(\ref{2+2EM}).

Consequently, for any QG theory with $2\alpha+\beta\neq0$ that admits a DP geometry containing 2D factors, a constant EM field component associated with each such factor is allowed. The restriction to $2\alpha+\beta\neq0$ is essential here, since for $2\alpha+\beta=0$ the existence of the KVFs is not guaranteed, and the most general form of the metric is therefore not known.

\section{Graphical summary of the solution classes}
\label{Agraph}
%%%%%%%%%%%%%%%%%%%%%%%%%%%%%%

In this Appendix, we offer a tabular summary of the sets of solutions obtained in this paper for the two theories discussed here, LG and QG, with the Weyl-squared action ({denoted as $W^2$ in tables,} which reduces to WCG for $n=4$) serving as an explicit example of a particular choice of QG couplings. Five tables summarize these sets for the individual DP metric structures considered here (Tables~\ref{Apm+0}--\ref{Ap2+...+2}), and two further tables are devoted specifically to the 4D and 6D cases (Tables~\ref{Ap4D} and~\ref{Ap6D}).

The tables list the solutions (meaning, in our case, conditions on the Ricci scalars, the dimensions of the DP subspaces, and, for QG, the coupling constants) for each theory, together with the relation of the LG solutions to them (this entry is left blank for LG itself). For QG, if the entry describing the relation of the LG solutions to the QG ones says ``equal'' or ``subset'', this applies only to QG theories that are nontrivial (i.e., admit solutions other than Minkowski) within the considered DP structure.

Additional information about a particular solution is given in square brackets. The label ``implicit'' indicates that the corresponding conditions are not in the form of Ricci scalars expressed from the equations, but rather in the form of equations mixing these scalars, the dimensions, and/or the coupling constants. Information about the algebraic type of the metric structures discussed, and therefore of the solutions, is given in the table captions.

\begin{widetext}

%---------------------------------------------------------------
\begin{table}[H]
\centering
\small
\renewcommand{\arraystretch}{1.4}
\begin{tabular}{l l l}
\hline\hline
\textbf{Theory} & \textbf{Solutions} & \textbf{Relation of LG solutions to them} \\
\hline\hline
\textbf{LG} & all allowed & -- \\
\hline
\textbf{QG} & \parbox[t]{0.46\textwidth}{\raggedright for $m=4$ and (\ref{QGccsol}): all allowed, else: only Minkowski} & {\strut equal} \\
\hline
\textbf{$W^2$} & all allowed & {\strut equal} \\
\hline\hline
\end{tabular}
\caption{Solutions obtained within the $(m+0)$ metric structure of constant curvature [type $O$].}
\label{Apm+0}
\end{table}

%---------------------------------------------------------------

\begin{table}[H]
\centering
\small
\renewcommand{\arraystretch}{1.4}
\begin{tabular}{l l l}
\hline\hline
\textbf{Theory} & \textbf{Solutions} & \textbf{Relation of LG solutions to them} \\
\hline\hline
\textbf{LG} & only Minkowski & -- \\
\hline
\textbf{QG} & \parbox[t]{0.46\textwidth}{\raggedright for (\ref{QGccsol}): all allowed, else: only Minkowski} & {\strut subset} \\
\hline
\textbf{$W^2$} & all allowed & {\strut subset} \\
\hline\hline
\end{tabular}
\caption{Solutions obtained within the $(m+1)$ metric structure with constant curvature $m$D part [type $O$].}
\label{Apm+1}
\end{table}

%---------------------------------------------------------------
\begin{table}[H]
\centering
\small
\renewcommand{\arraystretch}{1.4}
\begin{tabular}{l l l}
\hline\hline
\textbf{Theory} & \textbf{Solutions} & \textbf{Relation of LG solutions to them} \\
\hline\hline
\textbf{LG} & (\ref{m1+m2EST}) & -- \\
\hline
\textbf{QG} & \parbox[t]{0.46\textwidth}{\raggedright (\ref{m1+m2cc}) [implicit]\\[0pt]
[explicit examples conditioned by (\ref{m1+m2merging}): (\ref{explLGoverlapA})--(\ref{explLGoverlapB})]}
& \parbox[t]{0.30\textwidth}{\raggedright {\strut subset} for (\ref{m1+m2merging}) only} \\
\hline
\textbf{$W^2$} & (\ref{CFagain}) & {\strut not a subset} \\
\hline\hline
\end{tabular}
\caption{Solutions obtained within the $(m_1+m_2)$ metric structure, with both subspaces of constant curvature [type $D$ except case (\ref{m1+m2CF}), which is type $O$].}
\label{Apm_1+m_2}
\end{table}

%---------------------------------------------------------------

\begin{table}[H]
\centering
\small
\renewcommand{\arraystretch}{1.4}
\begin{tabular}{l l l l}
\hline\hline
\textbf{Theory} & \textbf{Branch} & \textbf{Solutions} & \textbf{Relation of LG solutions to them} \\
\hline\hline
\textbf{LG} & & (\ref{m1+m2EST}) [for $m_1 \equiv m$, $m_2=2$] & -- \\
\hline
\textbf{QG} & $2\alpha+\beta\neq0$:
 & \parbox[t]{0.40\textwidth}{\raggedright $b_3=0$: merge with Tab.~\ref{Apm_1+m_2} [for $m_1 \equiv m$, $m_2=2$]\\
   $b_3\neq0$: (\ref{2+m1st}) or (\ref{2+m2nd})}
 & \parbox[t]{0.28\textwidth}{\raggedright {\strut subset} for (\ref{QGm+2LGasub1}) and (\ref{QGm+2LGasub2}), or when merging with Tab.~\ref{Apm_1+m_2}} \\
\cline{2-4}
 & $2\alpha+\beta=0$:
 & \parbox[t]{0.40\textwidth}{\raggedright (\ref{boxrce}) or (\ref{boxrce3}) or (\ref{onlyover}) or (\ref{boxrce2a})--(\ref{boxrce2b})
   [last two differential -- solution examples in App.~\ref{Abox} and App.~\ref{Afullbox}, respectively]}
 & {\strut not a subset} \\
 \hline
 \textbf{$W^2$} & & (\ref{CFagain}) [for $m_1 \equiv m$, $m_2=2$] & {\strut not a subset} \\
\hline\hline
\end{tabular}
\caption{Solutions obtained within the $(m+2)$ metric structure with $m$D part of constant curvature and $m \neq 2$ [type $D$ except case (\ref{2+mCF}), which is type $O$].}
\label{Apm+2}
\end{table}

%---------------------------------------------------------------

\begin{table}[H]
\centering
\small
\renewcommand{\arraystretch}{1.4}
\begin{tabular}{l l l l}
\hline\hline
\textbf{Theory} & \textbf{Branch} & \textbf{Solutions} & \textbf{Relation of LG solutions to them} \\
\hline\hline
\textbf{LG} & & (\ref{2+...+2EST}) & -- \\
\hline
\textbf{QG} & $2\alpha+\beta\neq0$:
 & \parbox[t]{0.40\textwidth}{\raggedright (\ref{2+...b3k=0}) [implicit] or (\ref{2+...+2cond1}) [alternative form: (\ref{inbs})]\\[0pt]
   [example conditioned by (\ref{2+...+2LGsubstandard}): (\ref{LGinQG2+...+2})]}
 & \parbox[t]{0.28\textwidth}{\raggedright {\strut subset} for $n=4$ or (\ref{2+...+2LGsubstandard})} \\
\cline{2-4}
 & $2\alpha+\beta=0$:
 & \parbox[t]{0.40\textwidth}{\raggedright (\ref{2+...+2gna1}) or (\ref{2+...+2g=a}) or (\ref{2+...+2gna2})
   [last two differential \\ -- solution examples in App.~\ref{Abox}]}
 &{\strut subset} for $n=4$ or $\alpha=-\gamma$ \\
 \hline
\textbf{$W^2$} & & (\ref{2+...b3k=0}) [for (\ref{coeffCG})] & {\strut (generally) not a subset} \\
\hline\hline
\end{tabular}
\caption{Solutions obtained within the $(2+...+2)$ metric structure [type $D$ except the $(2+2)$ case with $R_{(1)}=-R_{(2)}$, which is type $O$].}
\label{Ap2+...+2}
\end{table}

%---------------------------------------------------------------
\begin{table}[H]
\centering
\small
\renewcommand{\arraystretch}{1.4}
\begin{tabular}{l l l l}
\hline\hline
\textbf{Theory} & \textbf{Structure} & \textbf{Solutions} & \textbf{Relation of LG solutions to them} \\
\hline\hline
\textbf{LG} & $(4+0)$ & all allowed & -- \\
\cline{2-4}
 & $(3+1)$ & only Minkowski & -- \\
\cline{2-4}
 & $(2+2)$ & $R_{(1)}=R_{(2)}=2\Lambda$ & -- \\
\hline
\textbf{QG} & $(4+0)$ & all allowed & {\strut equal} \\
\cline{2-4}
 & $(3+1)$ & \parbox[t]{0.40\textwidth}{\raggedright for $\beta=-3\alpha$: all allowed, else: only Minkowski} & {\strut subset} \\
\cline{2-4}
 & $(2+2)$, $2\alpha+\beta\neq0$: & (\ref{4D2+2full}) [$\beta=-3\alpha$] or (\ref{4D2+2bezb3}) & {\strut subset} \\
\cline{2-4}
 & $(2+2)$, $2\alpha+\beta=0$: & \parbox[t]{0.40\textwidth}{\raggedright (\ref{2+2box}) [differential -- solution examples in App.~\ref{Abox}]} & {\strut subset} \\
 \hline
\textbf{$W^2$} & $(4+0)$ & all allowed & {\strut equal} \\
\cline{2-4}
 & $(3+1)$ & all allowed & {\strut subset} \\
\cline{2-4}
 & $(2+2)$ & (\ref{4D2+2full}), see~\cite{previous} & {\strut subset} \\
\hline\hline
\end{tabular}
\caption{Solutions obtained in 4D within the metric structures considered [type $O$ for $(4+0)$, $(3+1)$ and the case $R_{(1)}=-R_{(2)}$ of $(2+2)$, elsewhere type $D$].}
\label{Ap4D}
\end{table}

%---------------------------------------------------------------

%---------------------------------------------------------------
\begin{table}[H]
\centering
\small
\renewcommand{\arraystretch}{1.4}
\begin{tabular}{l l l l}
\hline\hline
\textbf{Theory} & \textbf{Structure} & \textbf{Solutions} & \textbf{Relation of LG solutions to them} \\
\hline\hline
\textbf{LG} & $(6+0)$ & all allowed & -- \\
\cline{2-4}
 & $(5+1)$ & only Minkowski & -- \\
\cline{2-4}
 & $(4+2)$ & $R_{(4D)}=2\Lambda=2R_{(2D)}$ & -- \\
\cline{2-4}
 & $(3+3)$ & $R_{(1)}=R_{(2)}=3\Lambda/2$ & -- \\
\cline{2-4}
 & $(2+2+2)$ & $R_{(1)}=R_{(2)}=R_{(3)}=\Lambda$ & -- \\
\hline
\textbf{QG} & $(6+0)$ & (\ref{QG6+0/5+1c}) & {\strut equal} \\
\cline{2-4}
 & $(5+1)$ & (\ref{QG6+0/5+1c}) & {\strut subset} \\
\cline{2-4}
 & $(4+2)$, $2\alpha+\beta\neq0$: & (\ref{2+m1st}) & {\strut not a subset} \\
\cline{2-4}
 & $(4+2)$, $2\alpha+\beta=0$:
 & \parbox[t]{0.40\textwidth}{\raggedright (\ref{boxrce}) or (\ref{onlyover}) [differential -- solution examples in App.~\ref{Abox}] or (\ref{boxrce2a})--(\ref{boxrce2b}) [for $m=4$, differential -- solution examples in App.~\ref{Afullbox}]}
 & {\strut not a subset} \\
\cline{2-4}
 & $(3+3)$ & (\ref{3+3cases}) & \parbox[t]{0.28\textwidth}{\raggedright {\strut subset} for 1) in (\ref{3+3cases}), {\strut equal} for case 2)} \\
\cline{2-4}
 & $(2+2+2)$ & see the two branches below & \parbox[t]{0.28\textwidth}{\raggedright generally {\strut subset} for (\ref{LGcontainedinQG}) theories} \\
 & \hspace{0.8em} $2\alpha+\beta\neq0$: & \parbox[t]{0.40\textwidth}{\raggedright (\ref{2+2+2a}) with $3\alpha+\beta=0$ or (\ref{6Dcc2+2+2})}
 & \parbox[t]{0.28\textwidth}{\raggedright {\strut equal} for 2) and {\strut subset} for 1) of (\ref{6Dcc2+2+2}), else {\strut not a subset}} \\
 & \hspace{0.8em} $2\alpha+\beta=0$: & (\ref{6Ddiff1}) or (\ref{6Ddiff2})
 & \parbox[t]{0.28\textwidth}{\raggedright {\strut subset} for (\ref{2+...+2g=a}) only} \\
 \hline
\textbf{$W^2$} & $(6+0)$ & all allowed & {\strut equal} \\
\cline{2-4}
 & $(5+1)$ & all allowed & {\strut subset} \\
\cline{2-4}
 & $(4+2)$ & $R_{(2D)}=-R_{(4D)}/6$ & {\strut not a subset} \\
\cline{2-4}
 & $(3+3)$ & $R_1=-R_2$ & {\strut not a subset} \\
\cline{2-4}
 & $(2+2+2)$ & only Minkowski & {\strut not a subset} \\
\hline\hline
\end{tabular}
\caption{Solutions obtained in 6D within the metric structures considered [type $O$ for $(6+0)$, $(5+1)$, the case $R_{(4D)}=-6R_{(2D)}$ of $(4+2)$ and the case $R_{(1)}=-R_{(2)}$ of $(3+3)$, elsewhere type $D$].}
\label{Ap6D}
\end{table}

\end{widetext}


\begin{thebibliography}{9}

\bibitem{previous} P. Jizba and T. Lehe\v{c}kov\'{a}, Phys. Rev. D \textbf{113} (2026) 044060.
%Generalized Birkhoff theorems and 2+2 direct pruduct spacetimes in Weyl conformal gravity

\bibitem{Bertotti1959}
B.~Bertotti, 
%Uniform electromagnetic field in the theory of general
%relativity, 
Phys. Rev. {\bf 116} (1959) 1331.

\bibitem{Robinson1959}
I.~Robinson, 
%A solution of the Maxwell-Einstein equations, 
Bull. Acad.
Polon. Sci. Ser. Sci. Math. Astr. Phys. {\bf 7} (1959) 351.

\bibitem{KunduriLucietti2013}
H.K.~Kunduri and J.~Lucietti, 
%Classification of near-horizon geometries
%of extremal black holes, 
Living Rev. Relativity {\bf 16} (2013) 8.

\bibitem{Maia1985}
M.D.~Maia, 
%Geometry of Kaluza-Klein theory. I. Basic setting, 
Phys. Rev. D {\bf 31} (1985) 262.

\bibitem{OverduinWesson1997}
J.M.~Overduin and P.S.~Wesson, 
%Kaluza-Klein gravity, 
Phys. Rept. {\bf 283} (1997) 303.


\bibitem{EmparanReall2008}
R.~Emparan and H.~S.~Reall, 
%Black Holes in Higher Dimensions,
Living Rev. Relativity \textbf{11} (2008) 6.


\bibitem{Stelle1977}
K.S.~Stelle, 
%Renormalization of Higher-Derivative Quantum Gravity,
Phys. Rev. D {\bf 16} (1977) 953.

\bibitem{Stelle1978}
K.S.~Stelle, 
%Classical gravity with higher derivatives, 
Gen. Rel. Grav. {\bf 9} (1978) 353.

\bibitem{PravdaEtAl2017}
V.~Pravda, A.~Pravdov\'{a}, J.~Podolsk\'{y} and R. \v{S}varc, 
%Exact solutions to quadratic gravity, 
Phys. Rev. D {\bf 95} (2017) 084025.

\bibitem{QG2} T. M\'{a}lek and V. Pravda, Phys. \ Rev. \ D \textbf{84} (2011) 024047.
%Type III and N solutions to quadratic gravity



\bibitem{QG3} I.~Gullu et. al., Phys. Rev. D  \textbf{83} (2011) 084015.

\bibitem{QG4} V.~Pravda, A.~Pravdov\'{a} and G.~Turner, Phys. Rev. D \textbf{110} (2024) 044069.

\bibitem{QG5} R.~\v{S}vanc, J. Podolsk\'{y}, V. Pravda and A.~Pravdov\'{a}, Phys. Rev. Lett. \textbf{121}, 231104.


\bibitem{LuPerkinsPopeStelle2015a}
H.~L\"u, A.~Perkins, C.~N.~Pope, and K.~S.~Stelle, 
%Black Holes in Higher-Derivative Gravity, 
Phys. Rev. Lett. \textbf{114} (2015) 171601.

\bibitem{LuPerkinsPopeStelle2015b}
H.~L\"u, A.~Perkins, C.~N.~Pope, and K.~S.~Stelle, 
%Spherically Symmetric Solutions in Higher-Derivative Gravity, 
Phys. Rev. D \textbf{92} (2015) 124019.

\bibitem{Bach1921}
R.~Bach, 
%Zur Weylschen Relativitätstheorie und der Weylschen Erweiterung des Krümmungstensorbegriffs, 
Math. Z. {\bf 9} (1921) 110.


\bibitem{Tangherlini1963}
F.R.~Tangherlini,
%``Schwarzschild field in $n$ dimensions and the dimensionality of space problem,''
Nuovo Cimento \textbf{27} (1963) 636.


\bibitem{MyersPerry1986}
R.C.~Myers and M.J.~Perry,
%``Black holes in higher dimensional space-times,''
Ann. Phys. (N.Y.) \textbf{172} (1986) 304.


\bibitem{OrtaggioPodolsky2002}
M.~Ortaggio and J.~Podolsk\'{y}, 
%Impulsive waves in electrovac direct product spacetimes with Λ, 
Class. Quantum Grav. {\bf 19} (2002) 5221.

\bibitem{Kadlecova2009}
H.~Kadlecov\'{a}, A.~Zelnikov, P.~Krtou\v{s} and J.~Podolsk\'{y}, 
%Gyratons on direct-product spacetimes, 
Phys. Rev. D {\bf 80} (2009) 024004.

\bibitem{Podolsky2012}
P.~Krtou\v{s}, J.~Podolsk\'{y}, A.~Zelnikov and H.~Kadlecov\'{a},
%Higher-dimensional Kundt waves and gyratons, 
Phys. Rev. D {\bf 86} (2012) 044039.

\bibitem{Pravda2007}
V.~Pravda, A.~Pravdov\'{a}, and M.~Ortaggio, 
%Type D Einstein spacetimes in higher dimensions, 
Class. Quantum Grav. {\bf 24} (2007) 4407.

\bibitem{Pravda2013}
M.~Ortaggio, V.~Pravda and A.~Pravdov\'{a}, 
%Algebraic classification of higher dimensional spacetimes based on null alignment, 
Class. Quantum Grav. {\bf 30} (2013) 013001.


\bibitem{GursesSismanTekin2012}
M.~G\"{u}rses, T.Ç.~Şişman and B.~Tekin, 
%New Exact Solutions of Quadratic Curvature Gravity, 
Phys. Rev. D {\bf 86} (2012) 024009.

\bibitem{book2} D. Lovelock and H. Rund, {\em Tensors, differential forms and variational principles}, (Dover Publications Inc., 1975).


\bibitem{Chern:1955}
S.-S. Chern, 
%"An elementary proof of the existence of isothermal parameters on a surface," 
Proc. Amer. Math. Soc. \textbf{6} (1955) 771.

\bibitem{Stephani:2003tm}
H.~Stephani, D.~Kramer, M.A.H.~MacCallum, C.~Hoenselaers, and E.~Herlt, {\em Exact Solutions of Einstein's Field Equations, 2nd ed.}, (Cambridge University Press, Cambridge, England, 2003).

\bibitem{STbook} J.B. Griffiths and J. Podolský, {\em Exact space-times in Einstein General relativity}, (Cambridge University Press, Cambridge, 2012).

\bibitem{Weinbergbook} 
S.~Weinberg, {\em Gravitation and Cosmology: Principles and Applications of the General Theory of Relativity}, (John Wiley, New York, 1972).

\bibitem{Weylnclass} 
A.~Coley, R.~Milson, V.~Pravda and A.~Pravdová, Class. \ Quant. 
Grav. \textbf{21} (2004) L35.
%Classification of the Weyl tensor in higher dimensions

\bibitem{QG1} I. Gullu and B. Tekin, Phys. \ Rev. \ D \textbf{80} (2009) 064033.
%Massive higher derivative gravity in D-dimensional anti-de Sitter spacetimes


\bibitem{Schmidt} V.D. Dzhunushaliev and H.J. Schmidt, J. \ Math. \ Phys. \textbf{41} (2000) 3007.
%New vacuum solutions of conformal Weyl gravity

\bibitem{KinnersleyWalker1970}
W.~Kinnersley and M.~Walker,
%``Uniformly accelerating charged mass in general relativity,''
Phys.\ Rev.\ D \textbf{2} (1970) 1359.

\bibitem{PlebanskiDemianski1976}
J.F.~Pleba\'nski and M.~Demia\'nski,
%``Rotating, charged, and uniformly accelerating mass in general relativity,''
Ann.\ Phys.\ (N.Y.) \textbf{98} (1976) 98.

\bibitem{KuhnelRademacher2016}
W.~K\"uhnel and H.-B.~Rademacher,
%``Conformally Einstein product spaces,''
Differ.\ Geom.\ Appl.\ \textbf{49} (2016) 65.


%\bibitem{DiasLemos2003}
%O.J.C.~Dias and J.P.S.~Lemos,
%``The extremal limits of the C-metric: Nariai, Bertotti--Robinson and anti-Nariai C-metrics,''
%arXiv:hep-th/0306194.

\bibitem{Petersen:16}
P.~Petersen, {\em Riemannian Geometry, 3rd ed.}, Graduate Texts in Mathematics 171, (Springer, New York, 2016).

\bibitem{Teitelboim1983}
C. Teitelboim, Phys. Lett. B \textbf{126} (1983)  41.

\bibitem{Jackiw1985}
R. Jackiw, Nucl. Phys. B \textbf{252} (1985) 343.

\bibitem{GKV2002}
D. Grumiller, W. Kummer, and D. V. Vassilevich, Phys. Rep. \textbf{369} (2002) 327.

\bibitem{HJS2D} H.-J. Schmidt, Gen. Rel. Grav. 31 (1999).
%`The classical solutions of two-dimensional gravity


\bibitem{Lu2013} H. L\"{u}, Y. Pang, and C.N. Pope, Phys. \ Rev. \ D \textbf{87} (2013) 104013. 


\bibitem{Anastasiou2021}
G.~Anastasiou, I.J.~Araya, and R.~Olea,
%``Einstein gravity from conformal gravity in 6D,''
J. High Energy Phys.  \textbf{01} (2021) 134.

\bibitem{BoulangerErdmenger2004}
N.~Boulanger and J.~Erdmenger,
%``A classification of local Weyl invariants in D=8,''
Class.\ Quantum Grav.\ \textbf{21} (2004) 4305.


\bibitem{BoulangerRovere2025}
N.~Boulanger and D.~Rovere,
%``8D conformal gravity with Einstein sector, and its relation to the $Q$-curvature,''
arXiv:2511.01368 [hep-th] (2025).

\bibitem{Ndimconf} A. Hell and D. L\"{u}st, J. High Energy Phys. \textbf{2025} (2025) 202.
%Conformal and pure scale-invariant gravities in d dimensions



\bibitem{Weylnclass2} 
M.~Ortaggio, V. Pravda and A. Pravdová, Class. \ Quant. \ Grav. \textbf{30} (2013) 013001.


\bibitem{Straumann:2013}
N.~Straumann, {\em General Relativity, Second Edition}, (Springer, Heidelberg, 2013).

%%%%%%%%%%%%%%%%%%%%%%%%%%%%%%%%%%%%%%%%%%%%%%%%%%%%%%%%%%%%%%%%%%%%%%%%%%%%%%%%%%%%%%%%%%%%


%\bibitem{MK+} P.D. Mannheim and D. Kazanas, Phys. Rev D \textbf{44} (1991) 417.
%Solutions to the Reissner-Nordström, Kerr, and Kerr-Newman problems in fourth-order conformal Weyl gravity

%\bibitem{topBHs} D.~Klemm, Class. Quant. Grav. \textbf{15} (1998) 3195.
%Topological black holes in Weyl conformal gravity


%\bibitem{Riegert} R.J. Riegert, Phys. Rev. Lett. \textbf{53} (1984) 315.
%Birkhoff's Theorem in Conformal Gravity



%\bibitem{Bekenstein1974} J.D.~Bekenstein, Ann.\ Phys.\ (N.Y.) \textbf{82} (1974) 535.
%``Exact solutions of Einstein-conformal scalar equations,''


%\bibitem{XanthopoulosZannias1989} B.C.~Xanthopoulos and T.~Zannias, Phys.\ Rev.\ D \textbf{40} (1989) 2564.
%``Einstein gravity coupled to a massless conformally invariant scalar field in arbitrary dimensions,''


%\bibitem{HassaineMartinez2007} M.~Hassa\"{\i}ne and C.~Mart\'{\i}nez,Phys.\ Rev.\ D \textbf{75} (2007) 027502.
%``Higher-dimensional black holes with a conformally invariant Maxwell source,''


%\bibitem{NegativeWeyl} H.J. Schmidt, Gen. Rel. Grav. \textbf{35} (2003) 937-938.
%The Square of the Weyl tensor can be negative


%\bibitem{Mannheim:92} P.D.~Mannheim and D.~Kazanas, Astrophys. J. \textbf{391} (1992) 429.
%“Exact vacuum solution to conformal Weyl gravity and galactic rotation curves,” 


%\bibitem{LJ:25} P.~Jizba and T.~Lehe\v{c}kov\'{a}, Eur. Phys. J. C \textbf{85} (2025) 602.
%Generalized Eddington–Finkelstein coordinates and exact Vaidya-type solutions in Weyl conformal gravity. 

%%%%%%

%\bibitem{DJT1982}S.~Deser, R.~Jackiw, and S.~Templeton, Ann. Phys. (N.Y.) \textbf{140} (1982) 372.

%\bibitem{HorneWitten1989}J.H.~Horne and E.~Witten, Phys. Rev. Lett. \textbf{62} (1989) 501.

%\bibitem{Lu2011} H.~L\"u, Y.~Pang, and C.N.~Pope, Phys.\ Rev.\ D \textbf{84} (2011) 064001.
%``Conformal gravity and extensions of critical gravity,''



\end{thebibliography}
\end{document}